\documentclass[twocolumn]{aastex63}

\shorttitle{Dust substructures triggered by two migrating super-Earths}
\shortauthors{Cui and Szuszkiewicz}

\begin{document}

\title{On the dust substructures triggered by two super-Earths migrating 
in low-viscosity disks}

\correspondingauthor{Zijia Cui}
\email{zijiacui@whpu.edu.cn}

\author[0000-0002-6222-5913]{Zijia Cui}
\affiliation{School of Electrical and Electronic Engineering, Wuhan Polytechnic University \\
Wuhan 430048, China}

\author[0000-0002-7881-2805]{Ewa Szuszkiewicz}
\affiliation{Institute of Physics and CASA*, University of Szczecin \\
Wielkopolska 15, PL-70-451, Szczecin, Poland}



\begin{abstract}

We investigate dust substructure formation induced by
two super-Earths migrating in a low-viscosity disk with single-size dust
grains selected from the submillimeter to centimeter range of sizes.
The orbital evolution of planets takes place in the vicinity of the 2:1 
commensurability, which allows to determine, in addition to the
dust substructure properties, the dust impact on the rate of
migration, the resonance capture, the libration overstability and   
the outcome of passage through the commensurability.
Using two-dimensional two-fluid hydrodynamical simulations with dust feedback
and dust diffusion taken into account, we identify  
two specific regions in the disk where the accumulation 
of dust particles is significant, leading to dust substructure formation
with the dust-to-gas ratio values close to or even higher than 1 for large grains.
The first region, with a narrow dust ring, 
is located between the planetary 
orbits and the second one,
with a broad feature, evolving in time in a multiple ring substructure,
is situated outside the orbit of the outer planet. Our results indicate that 
these two locations are favorable for planetesimal formation.
We discuss the properties of the dust substructures formed in our simulations
and outline possible consequences of their evolution for the
observed architectures of multi-planetary systems.

\end{abstract}



\section{Introduction} \label{sec:intro}

Disks of gas and dust rotating around newly born stars are the preferred  
sites of planet formation. High-resolution observations of such disks
in the optical/IR scattered-light by the Very Large Telescope (VLT) \citep[e.g.][]{2024AA..685A..53G},
Gemini \citep[e.g.][]{2022AJ...164A.109R}, and Subaru \citep[e.g.][]{2012ApJ...758..19H} as well as in the dust continuum and gas line emission by the Atacama Large Millimeter/submillimeter Array (ALMA)
\citep[e.g.][]{2018ApJ...869L.41A}
and the Very Large Array (VLA) \citep[e.g.][]{2025AA..694A..290G}
enable detection of the manifestations of processes taking place
during the early phases of planetary system formation and evolution.
In the context of our study, 
mapping the distribution of dust particles  
of sub-mm to cm sizes by ALMA and VLA is of particular relevance. 
The observations show characteristic disk substructures in a
large number of disks \citep{2023ASPC..534..605B}.
In the present investigation, we address the formation of such substructures 
and the consequences of their evolution in protoplanetary disks for planet formation. 

The substructures have a form of  
axisymmetric gaps, rings and central cavities 
as well as asymmetric features like spirals and crescents or arcs
\citep{2015ApJ...808L...3A, [and references therein] 2020ARAA...58.483A}.
Their origin is still being explored and a variety of
mechanisms have been proposed \citep[see][]{2023ASPC..534..423B}. 
One of the possible explanations invokes the presence of the (proto)planets
embedded in the disks and we adopt this particular scenario in our 
study. 

Once a planet is able 
to curve a partial gap in a disk, due to the planet-disk interactions, 
the gas pressure bumps are 
generated at the edges 
of the gap and trap the inward drifting dust particles to form the 
rings and gaps in the disk
\citep{2004A&A...425L.9P, 2006A&A...453L.1129P, 2015ApJ...809..93D, 
2018ApJ...868...48K, 2024ARA&A..62..157B}.  
Most of the studies consider the giant planet-disk interactions, but 
the low-mass planets in the disks with a very low level of viscous turbulence
can also curve partial gaps and form the pressure bumps \citep{2023ASPC..534..685P}.  

The rings with the significant enhancement of dust concentration, 
formed in the process of planet-disk interactions,
are preferred
places for planetesimal formation by streaming 
instability \citep{2005ApJ...620..459Y} 
or direct gravitational collapse \citep{1973ApJ...183.1051G, 2002ApJ...580..494Y},
 leading in turn to the formation of planets.
In consequence, the first generation of planets, may create
the second generation at the positions of enhanced concentrations
of dust.
In this scenario, the planet-induced dust substructures 
can be used  
to make an attempt to connect the early stages of the
planetary system formation to the later times of 
the evolution, when the protoplanetary
disk of gas and dust has mostly dissipated
and all the planets are already fully formed.

The observational features, which might be particularly helpful in making 
this connection, are the mean-motion commensurabilities or resonances (MMRs), 
which provide a powerful diagnostics of a migration history of the systems. 
The resonances can be formed in the process of 
convergent migration of two planets as a consequence 
of planet-disk interactions. So the understanding of the orbital migration 
in the disks of gas and dust is
crucial. 

The impact of the dust on the low-mass planets migration 
has been extensively studied varying 
the particle Stokes numbers, the dust-to-gas mass ratios, 
the stellar mass-accretion rates, the disk viscosity and
the pebble accretion
\citep{2023ApJ...953...97G, 2025ApJ...986..199G}; considering  
the effects of the development of small-scale, intense dust vortices 
\citep{2020MNRAS...497..2425H}, dust-void and dust filament structures
\citep{2018ApJ...855L..28B, 2020MNRAS.497.5540R, 2024A2526A...690A..41C, 2025AA...694A.279R, 2025AA...698A..21C}. 
The dust particles present in the disk, affecting 
the orbital migration of the planets, can  
influence formation of the mean-motion resonances.
For example, if the dust-to-gas ratio in the dust rings 
is sufficiently 
high, the resonance might be broken, providing an additional explanation
for the observed period ratio 
distribution near the MMRs in Kepler planet pairs
\citep{2011ApJS..197....8L, 2014ApJ...790..146F, 2015MNRAS.448.1956S}. 
However, also the resonance interaction between planets
can affect the distribution of dust in the disk. For instance,
if two low-mass planets 
are locked in a wide mean-motion resonance, like the 2:1 MMR, 
then the dust accumulation at the outer edge of the gap is more efficient  
and may lead to the formation of a bright ring, facilitating its
detection 
\citep{2020AA...641A.125M}. 
There are  numerous attempts to make a comparison between the 
substructure morphologies obtained in the simulations and those observed  
in the disks 
\citep[e.g.][]{2018ApJ...866..110D, 2017ApJ...843..127D, 2018ApJ...869L..46D, 2022AA...665A.122K, 2024AA...692A..45K}. 

In this study, we explore the dust substructure formation by 
two super-Earths migrating in an almost inviscid protoplanetary disk 
of gas and dust. In order to take advantage of the diagnostics offered 
by mean-motion resonances, we investigate the planets in the convergent
migration being close to the 2:1 commensurability. Moreover, we
choose the properties of the gaseous disk and planets in such a way 
to have a
temporary resonant capture with the clear occurrence of the libration
overstability in a purely gaseous case \citep{2014AJ...147..32G, 2024AA...686A.277A}. 
The dust component treated as a pressureless fluid with the initial dust-to-gas
ratio equal to 0.01 everywhere in the disk,  consists of single-size dust 
particles.  
We perform calculations
with five different dust grain sizes in the submillimeter-centimeter range.
Varying 
the size of dust particles, keeping the gaseous component unchanged allow us
to examine how the morphology of the dust substructures depend on the dust
grain size. Moreover, our calculations show how the dust affects the
orbital migration of planets, their capture into the 2:1 mean-motion 
resonance and further evolution
as well as the overstability present in the purely gaseous disk.

The results are obtained using two-dimensional 
two-fluid numerical
simulations including the aerodynamic drag, the feedback onto the gas
from the dust particles and dust diffusion. 
The dust growth and accretion on the planets have not been taken into account. 
The importance of allowing for the
turbulent dust diffusion and dust feedback on gas in this type of
calculations has been demonstrated in numerous works 
\citep[e.g.][]{2016AA...591A..86T, 2018ApJ...868...48K, 2018ApJ...854..153W,
2019ApJ...879L..19K, 2019ApJ...876..7S, 2020AA...635A..105P, 2023MNRAS...520..2913M}.

We identify two specific regions in the disk where the increase in 
dust particle concentration is significant, in particular, for
large grains. The first is located between the planetary
orbits and the second
is situated outside the orbit of the outer planet. Similar dust substructures
have been found also in the calculations by \cite{2025AA...703A.270R} 
in the case of two super-Earths migrating divergently away from the position
of the 2:1 resonance. 
For large dust grains, in our calculations,
the dust-to-gas ratio can easily exceed the value of 1, so the conditions 
present in those locations are
favorable for forming planetesimals via streaming instability
\citep[e.g.][]{2021AJ....161...96C}.
Both regions have already been identified in the literature as the promising 
sites for planet formation. The region between planets has been explored 
in the sandwiched scenario \citep{2024MNRAS.528.6538P}. 
In the present study, we show that
their findings hold also in the case of two low-mass planets migrating
in an almost inviscid disk. The region 
exterior to the orbit of the planet has been a subject of studies in
the context of the sequential scenario of planet formation \citep[e.g.][]{2024AA...688A..22L}.  
Our results indicate the possibility to extend this scenario also to
the case of low-mass planets. 

The hypothesis that a second generation of planets
can be formed
as an outcome of the 
follow-up evolution of the enhanced dust concentrations found 
in our calculations requires further study and it is not the subject
of this investigation. Instead, we performed 
a search through all
confirmed planetary systems to identify the observed 
configurations similar to those inferred from our calculations. We have 
found a few encouraging examples. One of those is TOI 1136  
with its well defined resonance structure \citep{2024AJ....167..70B}.
The intention of this search is just to
show an interesting pathway leading from the early stages of planet 
formation
to the configurations observed in the confirmed
planetary systems. We see this as a particular part of the bigger
picture discussed in \cite{2023ASPC..534..717D}. 
Future predictions coming from the numerical simulations
equipped with the high accuracy masses 
provided by the upcoming space mission PLATO, 
combined with the results from the  
new generations of ALMA and VLA as well as 
the forthcoming Square Kilometer Array (SKA), 
will allow to make a robust
link between different stages of planetary system evolution. 

The plan of this paper is as follows. 
In Section~\ref{sec:methods} we describe the numerical methods 
applied in this work. 
In Section~\ref{sec:single-case}, we present the results of 
our simulations for an isolated super-Earth migrating in a disk of gas and dust.
The main results of this work, namely the effects of dust dynamics
on the orbital evolution of 
two super-Earths 
embedded in a disk with a very low viscosity near the 2:1 mean-motion 
resonance  
are given in 
Section~\ref{sec:two-planets-v5}.
In Section~\ref{sec:two-planets-v3}, we show how different the outcome 
of the calculations performed in Section~\ref{sec:two-planets-v5} is,
if we adopt a significantly higher turbulent viscosity.  
In Section~\ref{sec:discussion}, we discuss the properties of
the dust substructures formed in our simulations, the
possible consequences for the  
observed architectures of  multi-planetary 
systems and the limitations of our study. 
The conclusions of this work are given in Section~\ref{sec:conclusion}.

\section{Disk model and numerical setup} \label{sec:methods}

We perform 2D hydrodynamical simulations of two interacting 
super-Earths with masses $m_i$, with i = 1 denoting the inner 
planet and i = 2 denoting the outer one, embedded in and 
interacting with a protoplanetary disk of gas and dust. 
The planets orbit a central star of mass $M_{\ast}$. For 
convenience we make use of the planet-to-star mass ratios 
$q_i = m_i/M_{\ast}$.   

\subsection{Basic equations}
\label{subsec:disk model}

In this work, we consider a geometrically thin protoplanetary disk,
for which we adopt a two-dimensional disk model together with 
a cylindrical polar coordinate system $(r, \phi)$ with its origin 
located at the central star. The star is treated as a point mass.
We consider only the case, when the dust grains are coupled to the 
gas, which means that the dynamics of the dust-gas interaction can 
be reproduced by the two-fluid approximation. Within this approximation    
the gas evolution is described by the continuity equation and the
equation of motion given as 
\begin{equation}\label{eq:gas-contimuum}
\frac{\partial \Sigma_{g}}{\partial t} + \nabla \cdot (\Sigma_{g} \vec{v}_{g}) = 0,
\end{equation}

\begin{equation}\label{eq:gas-motion}
\frac{\partial \vec{v}_{g}}{\partial t} + \vec{v}_{g} \cdot \nabla \vec{v}_{g} = - \frac{1}{\Sigma_{g}} \nabla P - \nabla \Phi + \vec{f}_{\nu} - \frac{\Sigma_{d}}{\Sigma_{g}} \frac{\vec{v}_{g} - \vec{v}_{d}}{t_{stop}},
\end{equation}
\noindent where $\vec{v}_{g}$ and $\vec{v}_{d}$ are the velocities
of the gas and dust respectively. The surface gas density is denoted
by $\Sigma_{g}$ and that of dust by $\Sigma_{d}$. 
$P$ is a vertically averaged gas pressure with $P = c_{s}^2\Sigma_{g}$ 
for a locally isothermal equation of state adopted in the simulations. 
Here $c_s$ is the isothermal sound speed. $\Phi$ is the gravitational 
potential while
$\vec{f}_{\nu}$ represents the viscous force per unit mass.  
The last term on the right hand side of Eq.(\ref{eq:gas-motion}) 
gives an acceleration due to drag between the gas and solid components.
The stopping time $t_{stop}$ measures the time for a solid particle 
to relax to the gas velocity.
In the Epstein regime, which is relevant for 
the dust particle sizes and the disk properties considered in this work 
it can be written as \citep{2005ApJ...623..482T}:
\begin{equation}\label{eq:t-stop}
t_{stop} = \frac{\pi s_{d} \rho_{p}}{2 \Sigma_{g} \Omega_{K}},
\end{equation}
\noindent where $s_d$ is the dust particle size, $\rho_{p}$ is the 
bulk density of the dust particles and $\Omega_{K}$ is the Keplerian 
angular velocity defined as $\Omega_{K} = \sqrt{GM_{\ast}/r^{3}}$, 
where $G$ is the gravitational constant.  
A convenient quantity 
to describe the properties of the dust particles is
the Stokes number $St$ related to
$t_{stop}$  as follows
\begin{equation}\label{eq:st-number}
St    
= t_{stop} \Omega_{K} = \frac{\pi}{2} \frac{s_{d} \rho_{p}}{\Sigma_{g}}.
\end{equation}

The dust particles in the disk are treated as a pressureless fluid, which
means that the dynamics of solids is assumed to follow the 
Navier-Stokes equations without the pressure forces, but with the drag forces 
coupling the gas and dust dynamics. Therefore, 
the equation of motion of the dust is written in the form:
\begin{equation}\label{eq:dust-motion}
\frac{\partial \vec{v}_{d}}{\partial t} + \vec{v}_{d} \cdot \nabla \vec{v}_{d} = - \nabla \Phi - \frac{\vec{v}_{d} - \vec{v}_{g}}{t_{stop}}.
\end{equation}
\noindent 

The continuity equation takes into account turbulent diffusion of the dust  
particles within the gas through a diffusive mass flux $\vec{j}$ and 
is given as:
\begin{equation}
\label{eq:dust-continumm}
\frac{\partial \Sigma_{d}}{\partial t} + \nabla \cdot (\Sigma_{d} \vec{v}_{d} + \vec{j}) = 0,
\end{equation}
\noindent where $\vec{j}$ is of the form
\begin{equation}
\label{eq:cof-1}
\vec{j} = - D (\Sigma_{g} + \Sigma_{d}) \nabla (\frac{\Sigma_{d}}{\Sigma_{g} + \Sigma_{d}}).
\end{equation}
\noindent For simplicity, we take the diffusion coefficient $D$ to be 
equal to the kinematic viscosity $\nu$ \citep{2023MNRAS...520..2913M, 2025AA...698A..21C}.

\subsection{Disk model}
\label{subsec:physical-disk}

The initial gas surface density profile is taken as
\begin{equation}
\Sigma_{g}(r) = \Sigma_{g,0}\left(\frac{r}{R_0}\right)^{-\sigma},
\end{equation}
where $\sigma = 1$ and $\Sigma_{g,0} = 1 \times 10^{-4}$ in units
of $M_{\ast}/R_0^2$ in 
all the simulations. This value of $\Sigma_{g,0}$ corresponds to 
$\Sigma_{g}$ = 33 g/cm$^{2}$ for $M_{\ast}$ = M$_{\odot}$ and 
$R_{0}$ = 5.2 au. 
This gives the disk mass less than the minimum mass solar nebula 
\citep{1981PThPs..70...35H}.
The disk is flared with a constant flaring index $f$, which gives 
the disk aspect ratio as 
\begin{equation}
h=\frac{H}{r} = h_{0}\left(\frac{r}{R_0}\right)^{f},
\end{equation}
where $H$ is the scale height of the disk. We apply $h_{0} = 0.05$ 
and $f = 0.25$ in all the simulations.

In each simulation the disk contains only one dust component 
and the dust particle size $s_{d}$ is fixed. 
For exploring the effects of dust dynamics generated by the dust 
particles with different sizes, we consider 
$s_{d}$ = 0.01, 0.1, 1, 2 and 4 cm while the bulk 
density of the dust is set to $\rho_{d}$ =2 g/cm$^{3}$ 
for all cases. The corresponding Stokes numbers are 
0.00095, 0.0095, 0.095, 0.19 and 0.38 respectively for 
$\Sigma_{g}$ = 33 g/cm$^{2}$ at $R_{0}$ = 5.2 au based 
on Eq.~(\ref{eq:st-number}). 

In order to achieve the steady-state background solution 
for the dust–gas interaction in our simulations with a constant 
particle size, we follow \cite{2022ApJ...936.93} in adopting 
the initial dust surface density profile and viscosity in the disk. 
The initial dust surface density profile is set as
\begin{equation}
\Sigma_{d} = \epsilon \Sigma_{g},
\end{equation}
where $\epsilon$ is the dust-to-gas ratio given 
in the form  
\begin{equation}
\label{eq:epsilon}
\epsilon = \frac{-(bd_{0}-a) - \sqrt{(bd_{0}-a)^{2}-4d_{0}^{2}c}}{2d_{0}},
\end{equation} 
where $d_{0} = \frac{a_{0}\epsilon_{0}}{\epsilon_{0}^{2}+b_{0}\epsilon_{0}+c_{0}}(r/R_{0})^{\sigma -1}$ 
with $a = -2\beta(\beta -1)v_{K} St$, 
$b = \beta(2\gamma + 3)$\footnote{\label{footnote1}The quantities $a$ an $b$
have been corrected for the typos present in \cite{2022ApJ...936.93}.},
$c = 2\beta^{2}(\gamma + 1)(1 + St^{2})$. Here, 
$\beta = \sqrt{1+h^{2}\frac{d{\log P}}{d{\log r}}}$, 
$v_K$ is the Keplerian velocity 
and $\gamma = \frac{d{\log \beta}}{d{\log r}} - \frac{1}{2}$.
The subscript "0" refers to the values at $r = R_{0}$
and $\epsilon_{0} = 0.01$. 
The turbulence in the disk is simulated by adopting the viscosity prescription 
$\nu = \alpha c_{s} H(r)$ \citep{1973AA...24..337S}, where the 
viscous parameter $\alpha$ is set following \cite{2022ApJ...936.93}, namely 
\begin{equation}
\label{eq:viscos-alpha}
\alpha = \alpha_{0} (r/R_{0})^{\sigma - 2f - 1}.
\end{equation}
The fiducial value of $\alpha_{0}$ in this work is taken to 
be $10^{-5}$, which correspond to a nearly inviscid disk. In some simulations
we also adopted $\alpha_{0} = 10^{-3}$ for comparison. 
The steady-state drift solutions
for each disk (without planets) with the different dust grain size $s_{d}$ 
have been obtained in 2000 orbits of the simulations 
(see Appendix~\ref{sec:a1} for more details).

\subsection{Numerical setup}
\label{subsec:setup}

The governing hydrodynamical equations are solved 
using the code FARGO3D \citep{2019ApJS..241..25}. 
The system of units are as follows: the mass unit is the mass of the 
central star $M_{\ast}$. The length unit is  
$R_{0}$ while the time unit is $2\pi(GM_{\ast}/R_0^3)^{-1/2}$. 
The use of the dimensionless units enables the results to be scaled 
so that they apply to different radii and corresponding initial surface 
densities. However, in presenting the results we adopted the particular 
parameters, namely $R_0 = 5.2$ au and $M_{\ast} = M_{\odot}$. Therefore,
the time unit is 11.86 yr.

We take the computational mesh with 1024 evenly spaced grids in the 
azimuthal direction and 700 logarithmically spaced grids in the radial 
direction while the computational domain extends from $r_{min} = 0.2R_0$ 
to $r_{max} = 7.0R_0$ in the radial direction and covers the whole 
$2\pi$ domain in azimuth. The adopted computational mesh and domain
are the same as in \cite{2024AA...686A.277A} to facilitate the
comparison of the planet evolution in the protoplanetary disks
with gas and dust, considered in this work, with the purely gaseous disk.
The standard outflow boundary conditions are applied at the disk 
boundaries. The wave killing-zones operate in the domains of 
$[0.2R_0, 0.268R_0]$ and $[6.32R_0, 7.0R_0]$ in the inner and outer boundaries, 
respectively \citep{2006MNRAS...370...529D}. 

The planets are placed on circular orbits in the steady-state
disk, described in the previous subsection.
Their masses start from zero and grow to their final masses 
during the first 10 orbits of the simulation. After that, 
the masses of the planets are fixed. Further accretion of the disk material
is not considered. 
When calculating the force per unit mass acting on the planet, the indirect 
term arising on account of the acceleration of the origin of the 
coordinate system is included. 
The gravitational potential $\Phi$ is the sum of the 
potential from the planets $\phi_{i}$ and the central star, where $i$ 
is taken to be 1 and 2 for denoting the inner and outer planet. 
The potential due to a planet with mass of $m_{i}$ has the form of:
\begin{equation}
\Phi_{i} = - \frac{Gm_{i}}{\sqrt{r^{2} - 2rr_{i}cos(\phi - \phi_{i}) + r_{i}^{2} + b^{2}r_{i}^{2}}},
\end{equation}
\noindent where $(r_{i}, \phi_{i})$ are the cylindrical 
coordinates of the planet and the softening parameter $b$ is taken 
to be $0.6h$  \citep{2012AA...541A..123M}.
The disk self-gravity can be 
neglected, as the surface densities of the gas used in our simulations 
are relatively low, corresponding to a Toomre $Q$ value of the order of 20.
However, for the consistent treatment of the disk-planet interaction, we
implement in all the simulations the torque correction given 
by \cite{2008ApJ...678....483}. 

\section{Migration of an isolated super-Earth in a nearly inviscid disk 
of gas and dust}
\label{sec:single-case}

We start our investigations from an isolated planet case embedded in 
a low-viscosity ($\alpha_0 = 10^{-5}$) gaseous disk containing dust 
particles of a single size $s_{d}$.  
The rationale for doing that is to facilitate the interpretation of the
migration of two interacting super-Earths in such a disk, studied in this work.
Accordingly, two single planet cases are considered.  
The first one corresponds to the presence of only the inner planet in the disk 
with the planet-to-star mass ratio $q = 1.5 \times 10^{-5}$ 
and its initial position $r_{p,0} = 1.23$.  
The second one, instead, is consistent with  the situation when the outer 
planet is present in the disk without the inner one. In this case
$q = 3.0 \times 10^{-5}$ and $r_{p,0} = 2.0$. In both cases planets are
initially in a circular orbit.

We run a series of 2D hydrodynamical simulations 
for each of those two cases 
in the gaseous disk with the dust particles 
of a given size $s_{d}$. The adopted $s_d$  are taken to be 
0.01, 0.1, 1, 2 and 4 cm, respectively. 
The Stokes numbers calculated from Eq.~(\ref{eq:st-number}) at the 
initial position of the planet in each case are given 
in Table~\ref{tab:st-num}. 
To explore the effects caused by the presence
of dust grains in the disk on the planet migration, we run also the 
simulations for a single planet migrating in a purely gaseous disk, 
hereafter called the `gas case'. 

\begin{table}[htbp]
\tablenum{1}
\centering
\label{tab:st-num}
\caption{Stokes numbers at the initial planet positions}
\vspace{3mm}
\begin{tabular}{l|lllll}
\hline
$s_{d}$ (cm)  & $0.01$ & $0.1$ & $1$ & $2$ & $4$ \\ 
\hline
$r_{p,0} = 1.23$ & $0.00117$ & $0.0117$ &$0.117$ & $0.234$ & $0.468$  \\
\hline
$r_{p,0} = 2.0 $ & $0.0019$ & $0.019$ &$0.19$ & $0.38$  & $0.76$    \\
\hline
\end{tabular}
\end{table} 

\begin{figure}[htbp]
\centering
\includegraphics[width=1.0\linewidth]{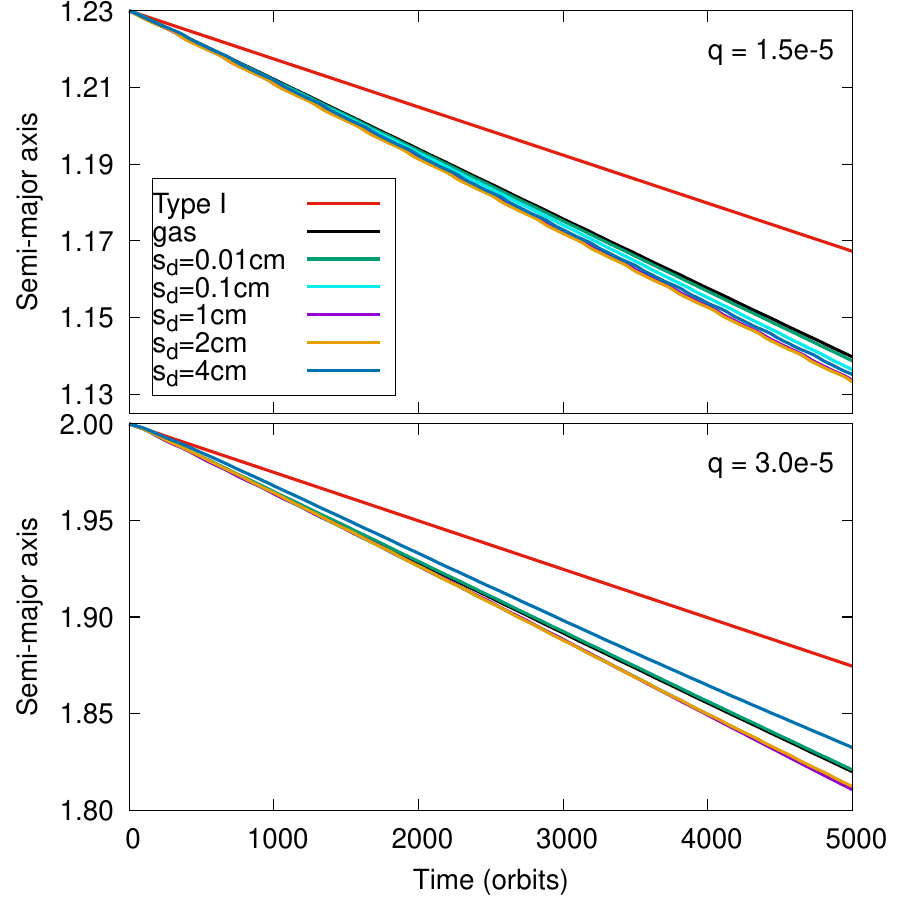}
\caption{Evolution of the semi-major axis of the planet
in a purely gaseous disk 
and in the disks of gas and dust with different $s_{d}$. 
The top and bottom 
panels show the results of the simulations for the planet of 
$q = 1.5 \times 10^{-5}$ with $r_{p,0} = 1.23$ and 
$q = 3 \times 10^{-5}$ with $r_{p,0} = 2.0$, respectively. 
The legend to the top panel applies also to the bottom one.
The red line in each panel represents the Type I migration calculated 
from Eq.~(\ref{eq:type1}).}
\label{fig:single-semi-evo}
\end{figure} 
 
In Figure~\ref{fig:single-semi-evo}, we show the evolution of the semi-major 
axis of a planet (lower-mass planet in the top panel; 
higher-mass planet in the bottom panel)
embedded in a disk of gas and dust with the grains 
of a single size $s_d$ for five different values of $s_d$, namely 
0.01 (green line), 0.1 (cyan line), 
1 (violet line), 2 (orange line) and 4 cm (dark blue line), 
during the first 5000 orbits of the simulations. 
Please note that in the bottom panel  
the orange line overlaps the violet one.
The migration of planets in the gas case (black line) 
and that predicted by assuming Type I migration (red line) are illustrated 
for comparison.
The Type I migration rate is calculated as follows
\citep{2002ApJ...565..1257}
\begin{equation}
\label{eq:type1}
\dot{r_{p}} = -2r_{p}\frac{\Gamma_{p}}{m_{p}\sqrt{GM_{\ast}r_{p}}},
\end{equation}
\noindent where $\Gamma_{p}$ is the Type I migration torque in 
a locally-isothermal disk with the form of \citep{2010MNRAS...401.1950}
\begin{eqnarray}
\frac{\Gamma_{p}}{\Gamma_{0}} = & - (2.5 - 0.5\tau - 0.1\sigma)(\frac{0.4}{b/h}) ^{0.71} \nonumber \\
                                & - 1.4\tau(\frac{0.4}{b/h}) ^{1.26} + 1.1(1.5 - \sigma)(\frac{0.4}{b/h})
\end{eqnarray}
with
\begin{equation}
\Gamma_{0} = (\frac{q}{h})^{2}\Sigma_{p}r_{p}^{4}\Omega_{p}^{2},
\end{equation}
\noindent where $\tau$ is the index of the disk temperature profile 
$T \propto r^{-\tau}$. For the disk parameters adopted in this work 
(Section~\ref{subsec:setup}) 
we have $\tau = 0.5$.  
In all simulations, the planets migrate inward with the rates larger 
than predicted by the classical Type I migration.

\begin{figure}[htbp]
\centering
\includegraphics[width=1.0\linewidth]{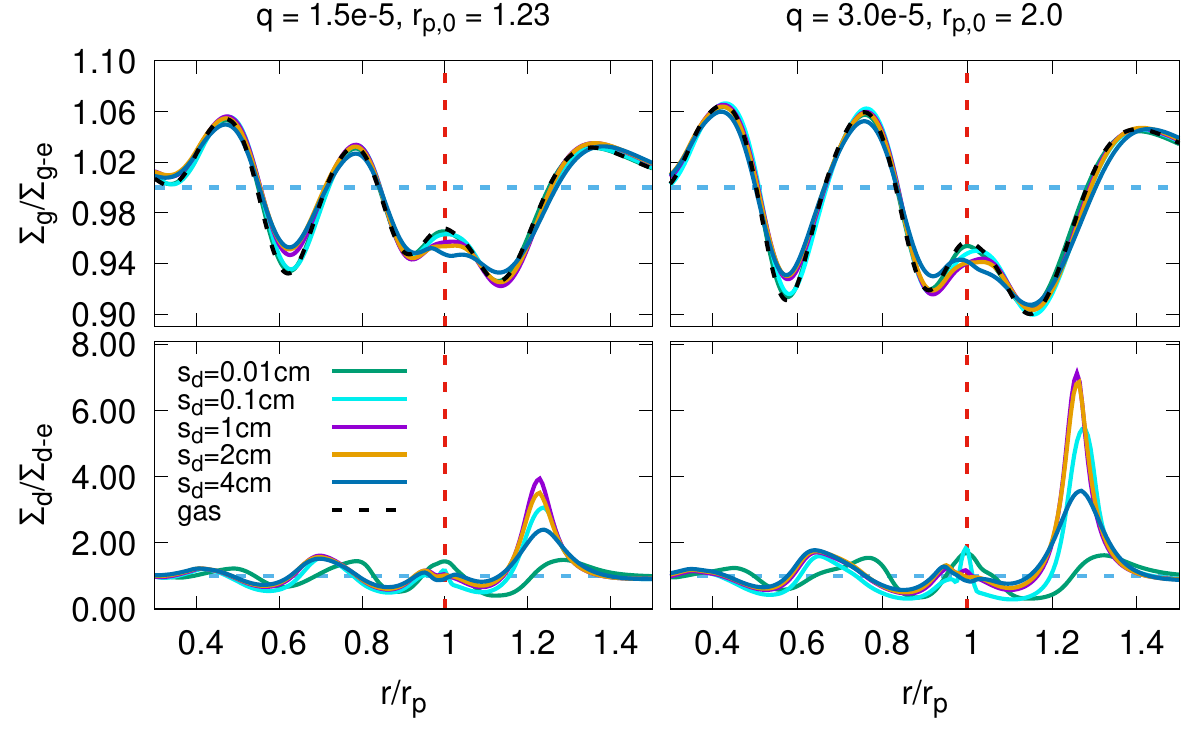}
\caption{The relevant gas and dust surface density $\Sigma_{g}/\Sigma_{g-e}$, 
$\Sigma_{d}/\Sigma_{d-e}$ around the planet at t = 5000 orbits in 
the simulations for a single planet with various $s_{d}$ and in the 
gas case. The description of lines are the same in both panels. 
The dashed red vertical line in each panel indicates the position 
of the planet. The dashed light blue horizontal line represents the unperturbed
background surface density.}
\label{fig:single-sden}
\end{figure}

The presence of dust grains in the disk selected for 
this study 
modifies the rates,
but does not reverse the direction of the migration. 
The migration rates depend on the changes of the disk structure caused by
the interaction of the disk material with the planet. 
Those changes can be observed 
in Figure~\ref{fig:single-sden}, where 
the azimuthally averaged gas (top panel) and dust (bottom panel) 
surface densities around the planet at t = 5000 orbits are presented. 
$\Sigma_{g-e}$ is the steady-state profile of $\Sigma_{g}$ obtained in 
the simulations for the empty disk at t = 2000 orbits.
$\Sigma_{d-e}$ and $\Sigma_{d}$ are analogous quantities for the dust.
The gas distribution in the vicinity of each planet forms 
a very shallow double gap. 
The presence of grains affects significantly
co-orbital region, which results in the differences of the migration rates
seen in Figure~\ref{fig:single-semi-evo}. 
It is interesting to notice also the appearance of 
a narrow partial gap formed in the region interior to the planet position,
namely at $r/r_{p} \sim 0.6$.     

In the dust distribution there are several rings present 
on both sides of the planet. Their locations corresponds to the local pressure 
gradient peaks related to the profile of the gas surface density
showed in the top panel in Figure~\ref{fig:single-sden}.  
In the case of the more massive planet ($q=3\times 10^{-5}$) in the disk with
the smallest dust grains $s_d=0.01$ cm, the dust rings are not very dense
and the maxima of their dust surface density are located  approximately 
at $r/r_{p} \sim 0.46$, 0.77 and 1.32. There is also a small overabundance 
of the dust particles in the co-orbital region ($r/r_{p}\sim 1)$. 
Interestingly enough,
the positions of the rings are close to (not in) the 1:3, 2:3 and 3:2 
resonances with the planet. 
For lower mass planet ($q=1.5\times 10^{-5}$) the locations of dust rings
are shifted slightly relative to those in the case of more massive one and
they are  at $r/r_{p} \sim 0.49$, 0.79 and 1.28.
In the disks with larger dust grains the rings are moved in the direction
of smaller $r/r_p$.  
The peak of the most pronounced dust ring is the highest in the disk with 
$s_{d} = 1$ cm grains and the lowest when $s_{d} = 0.01$ cm.
This is true for both planets.
The gas and dust features obtained in our simulations are qualitatively
similar to the results of \cite{2006A&A...453L.1129P}
 and \cite{2017ApJ...843..127D}. 

\begin{figure*}[htbp]
\centering
\includegraphics[width=1.0\linewidth]{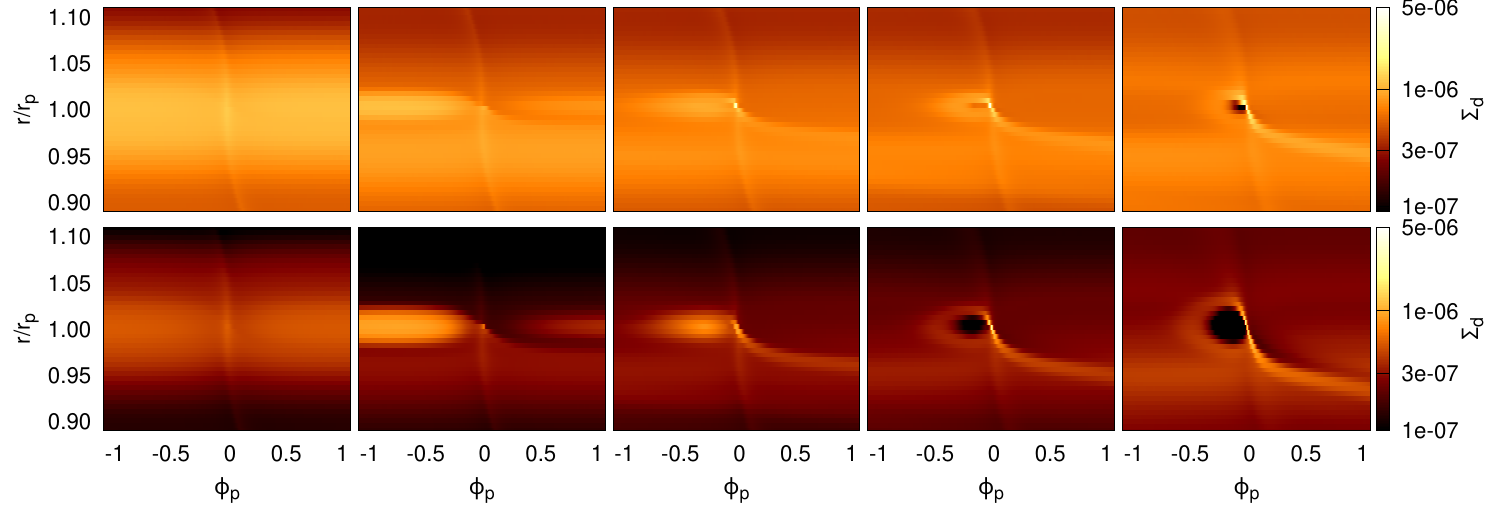}
\caption{From left to right: contour plots of $\Sigma_{d}$ in the vicinity of the planet 
in the disk with $s_{d}$ = 0.01 cm, 0.1 cm, 1 cm, 2 cm and 4 cm at t = 5000 orbits. 
Top panels show the cases of $q = 1.5 \times 10^{-5}$ with $r_{p,0} = 1.23$ while 
bottom panels show the cases of $q = 3.0 \times 10^{-5}$ with $r_{p,0} = 2.0$.
The planet is located in the center of each panel.}
\label{fig:single-dsden-contour}
\end{figure*} 

To complete our discussion of the effects of the dust presence in the
gaseous disk on the migration of 
an isolated super-Earth, we show 
in Figure~\ref{fig:single-dsden-contour} 
the contour plots of the dust
surface density $\Sigma_{d}$ in the vicinity of the planet at t = 5000 orbits 
in the simulations with various $s_{d}$.
The top panels show the results for $q = 1.5 \times 10^{-5}$ 
with $r_{p,0} = 1.23$ and the bottom panels are for 
$q = 3.0 \times 10^{-5}$ with $r_{p.0} = 2.0$. 
The dust particles with $s_{d} = 0.01$ cm
are accumulated in the 
corotation region of the planet, forming a co-orbital 
dust ring. If $s_{d}=0.1$ cm, there is a dust overdensity
situated behind the planet. This asymmetry is much more pronounced 
in the case of the more massive planet. Similar features has been found 
also in \cite{2019ApJ...884..178W} for a Neptune mass planet on 
a fixed orbit in the disk with $s_{d} = 0.1$ cm grains and $\alpha = 10^{-5}$. 
When $s_{d}$ is equal to 1 cm
this dust structure, as shown in our calculations, becomes 
smaller and a filament (high-density structure)
forms in front of the planet for both isolated planet cases.
For $s_{d} = 2$ cm, apart from the filament,
a small dust-void (low-density structure) appears behind the planet.
The size of the dust-void is bigger for the bigger mass planet 
and the larger dust grain size, as can 
be seen from the comparison between the $s_{d}=2$ and 
$s_{d}=4$ cm cases.
Moreover, the dust-void is surrounded by the high
density ring-shaped region. 

Previous studies show that these particular dust structures formed in the 
vicinity of the planet may produce a strong positive torque to reverse 
the planet migration \citep{2018ApJ...855L..28B, 2025AA...698A..21C}.
However, for our set of parameters describing the disk and planet properties 
this is not the case. 
As it is clear from Figure~\ref{fig:single-semi-evo}, in our simulations
the isolated planets 
migrate inward.
The total torques acting on them is always negative. 
The torque resulting from the dust component is positive only for
$q = 3.0 \times 10^{-5}$ with $r_{p,0} = 2.0$ and 
$s_{d}=4$ cm, but also in this case it is smaller than the negative 
torque coming from the gas component.

It is important to mention that the simulations of the isolated super-Earths
described in this section have been performed with the aim to set the
ground for the investigation of the 
migration of two interacting super-Earths in 
a disk of gas and dust.  
For this reason, we have considered here only a narrow range of 
parameters for the disk and planets relevant to our primary goal.  

\section{Orbital evolution of two super-Earths near the 2:1 resonance in 
a nearly inviscid disk of gas and dust}
\label{sec:two-planets-v5}

In this section, equipped with the knowledge about the orbital evolution of
an isolated planet in the protoplanetary disks of gas and dust, gained in 
Section~\ref{sec:single-case}, we move on to study the migration of two 
super-Earths embedded in such disks. 
The masses and the initial positions of the inner and outer planets are 
the same as those set up in Section~\ref{sec:single-case}, namely 
$q_{1} = 1.5 \times 10^{-5}$ with $r_{1,0} = 1.23$ 
and $q_{2} = 3 \times 10^{-5}$ with $r_{2,0} = 2.0$. 
Both planets are initially in circular orbits and start their evolution
in the vicinity of the 2:1 MMR. 
The aim is to investigate the effects that dust dynamics has on 
the rate of the planetary migration,  
attainment and stability of the 2:1 commensurability, as well as to
characterize the dust substructures triggered by the super-Earths during 
their evolution.

\subsection{Migration rates, excitation of the eccentricities and the
attainment of the 2:1 resonance}

\begin{figure*}[htbp]
\centering
\includegraphics[width=1.0\linewidth]{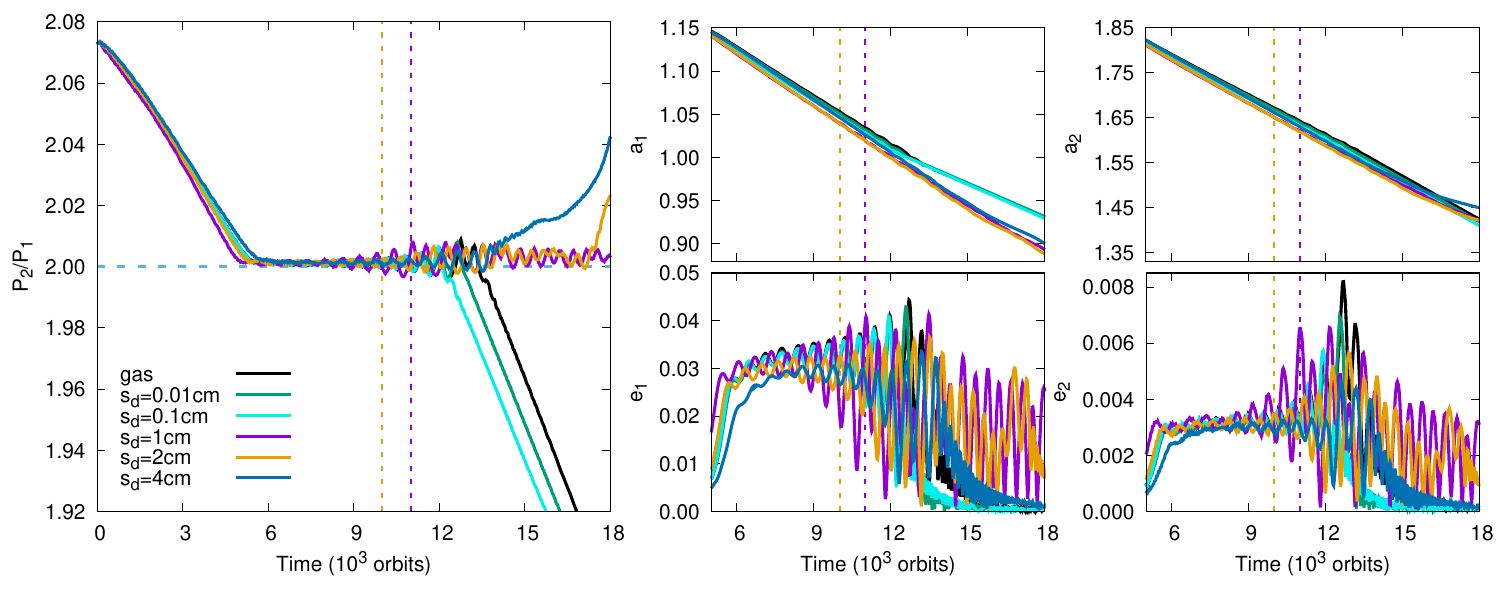}
\caption{From left to right: Evolution of the period ratios, semi-major axes 
and eccentricities of two planets in the purely gaseous disk and in the disks 
of gas and dust with different $s_{d}$. In the middle and right panels
we show the most relevant part of the evolution, omitting the first 5000 orbits
during which no significant differences can be seen. 
The viscous parameter $\alpha_{0}$ 
is taken to be $10^{-5}$. The horizontal dashed light blue line in the left
panel indicates the position of the 2:1 MMR.
The vertical dashed orange and violet lines separately indicate the moment of time when
the local dust-to-gas ratios $\epsilon \approx 1$ in the disks with the large grains ($s_d$ = 2 and 4 cm) and the moderate grains ($s_{d}$ = 1 cm). In the disks with small grains ($s_d$ = 0.01, 0.1 cm), $\epsilon$ is less than 1
till the end of our simulations. 
}
\label{fig:evo-two-v5}
\end{figure*}

In Figure~\ref{fig:evo-two-v5}, we show the results of the simulations 
for two planets migrating in the gaseous disks with the dust particles. 
The sizes of the particles are taken to be $s_d =0.01$ cm (green line),
0.1 cm (cyan line), 1 cm (violet line), 
2 cm (orange line) and 4 cm (dark blue line), respectively. For comparison, 
we present also the orbital evolution of 
planets in the gaseous disk without dust (black line).

In all considered here cases, the planets migrate inward for the whole 
duration of our calculations. Moreover, initially the migration is convergent 
and there are no significant differences in the migration rates  
in the disks with different dust grains. 
The convergent migration results in the occurrence of the 
2:1 MMR resonance. 
Once in the resonance, the migration is the slowest in the disk without dust for both planets.
In the disk with dust the migration rate increases along the grain size
sequence: $s_d$ =0.01, 0.1, 1 and 2 cm. However, 
for the disk with the biggest grains adopted here 
($s_d=4$ cm) the migration is slower than in the case of $s_d=2$ cm, 
but faster than in the case of $s_d=0.1$ cm.  

In the gaseous disk without dust, the planets arrive into the 2:1 MMR at
$t \sim 5300$ orbits. At that time the eccentricities are excited to 
$e_{1} = 0.03$ and $e_{2} = 0.003$ as shown in the middle panels. 
The characteristic overstability shown in detail in \cite{2024AA...686A.277A}
leads to the breaking of the resonance and 
at $t \sim 13300$ orbits the planets pass through the 2:1 MMR while 
both eccentricities begin to damp. At the end of the calculation, 
$e_{1}$ and $e_{2}$ are zero. The orbital evolution in this simulation 
confirm the results for the fiducial model in \cite{2024AA...686A.277A}. 

When dust particles are present in the disk, the orbital evolution of 
two planets differs from that in the gas case in two particular aspects.  
One of the main differences is the time during which the planets remain 
in the resonance and the second is the divergent migration, which occurs
only in the disks with the large grains ($s_{d} = 2$ and 4 cm) towards 
the end of our simulations. 
This can be clearly seen in the left panel 
of Figure~\ref{fig:evo-two-v5}. 
For the small grains with $s_d = 0.01$ cm and $s_d = 0.1$ cm the planets 
enter the resonance 2:1 at approximately the same time as in the gas case
($t \sim 5300$ orbits), but leave the resonance earlier, 
namely at about 12800 and 12400 orbits, respectively.
When the planets are locked in the 2:1 
resonance their eccentricities are excited to the similar values as 
in the gas case.

In the $s_{d}$ = 1 cm case, the planets enter 
the 2:1 MMR at $t \sim 4900$ orbits, slightly earlier than in other 
cases and then remain close to the  
2:1 commensurability till the end of the simulation. 
At the beginning of the resonance, both eccentricities increase to
the similar value 
as in the case of the purely gaseous disk, but after t $\sim$ 13000 orbits 
they start to decrease, showing large amplitude oscillations and continue
doing so till the end of the calculations. 

In the $s_{d} = 2$ cm case, the planets arrive in the 2:1 MMR
at $t \sim 5200$ orbits and stay there for 
about 4000 orbits longer than in the gas case. 
After that, the divergent 
migration occurs at $t \sim 17000$ orbits and then the planets leave 
the 2:1 MMR. In this case both planets continue their migration inward but 
the migration rate of the outer planet is slowed down since 
$t \sim 17000$ orbits. This leads to the divergent migration. The excited 
eccentricities of two planets are slightly lower than those in  
the disk without or with the small grains.  

In the $s_{d} = 4$ cm case, the planets arrive into the 2:1 MMR at 
$t \sim 5600$ orbits, a bit later than in the case with the smaller grains, 
and leave the resonance due to the divergent migration at $t \sim 14000$ orbits at 
the similar time as the planets in the gas case. After leaving the resonance,  
the inward migration of the outer planet
slows down and at $t \sim 16000$ orbits 
the migration rate decreased significantly. 
The values of the eccentricities of both planets are the smallest among all 
presented here cases. 

In the chosen here configuration of two super-Earths migrating in the
low viscosity gaseous disk, we observe a temporary capture
into the 2:1 MMR and the overstability described 
in \cite{2024AA...686A.277A}. 
The results shown in Figure~\ref{fig:evo-two-v5} indicate that the presence 
of dust in such a disk affects the duration of the MMR. 
The planets migrating
in the disks with the dust grains with the size less or equal to 0.1 cm
pass through the 2:1 MMR earlier for the larger grain size. 
Instead, for the disks with the dust grains with the 
size of 1, 2 and 4 cm, the planets can stay in the 2:1 MMR for a longer time or 
leave the 2:1 MMR due to divergent migration.
For larger grains from this
range of grain sizes, the planets leave the resonance sooner. 
Anticipating the description of the dust surface density evolution in our
calculations, we would like to
point out that in the disks with moderate grains ($s_d$ = 1 cm) and large
grains ($s_d$ =2 and 4 cm) 
the local dust-to-gas ratio becomes comparable to 1 at about 10000 
(large grains, vertical dashed orange line in Figure~\ref{fig:evo-two-v5}) 
and $\sim$ 11000 orbits (moderate grains, vertical violet line in 
Figure~\ref{fig:evo-two-v5}).

\subsection{Gas and dust surface density evolution  
in the vicinity of the planets} 
\label{subsec:sden-dsden-evo}

To gain a better insight into the effects of the dust particles present in
the disk on the formation and stability of the 2:1 resonance between planets
in the super-Earth mass range, as well as into the properties of the 
substructures induced by the planets in the disk, 
we discuss in this subsection the evolution of gas and dust surface densities 
in the vicinity of the planets during their migration. 

\subsubsection{Evolution of the gas surface density}

\begin{figure*}[htbp]
\centering
\includegraphics[width=1.0\linewidth]{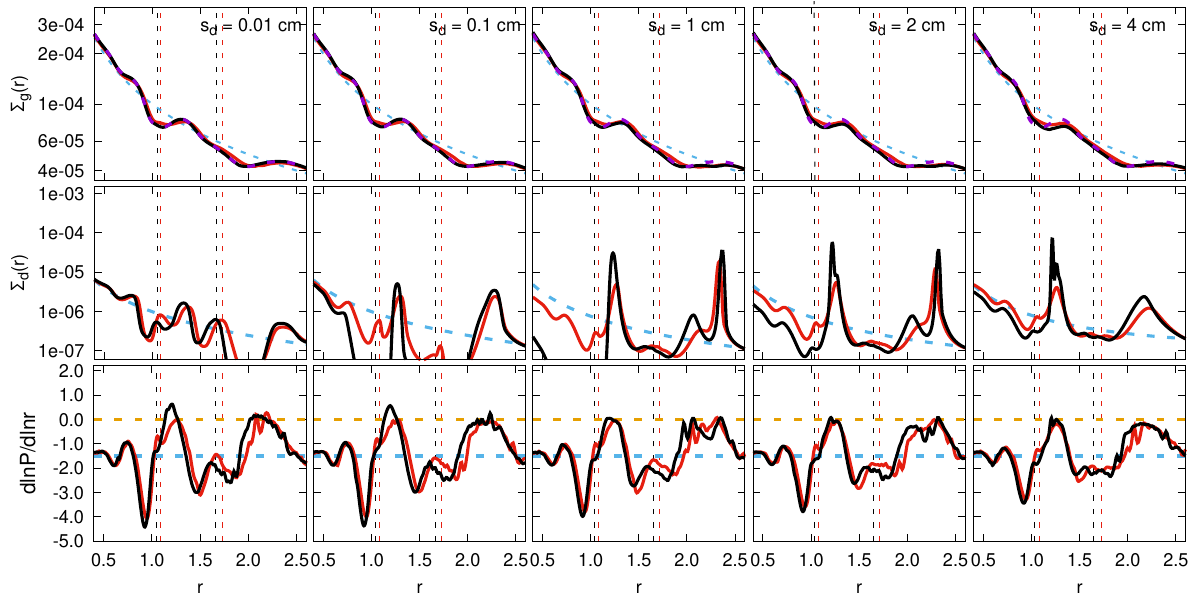}
\caption{Top and middle: azimuthally averaged gas and dust surface density 
as a function of $r$ at t = 8000 and 10000 orbits in the simulations 
with various $s_{d}$, which are indicated respectively by the red and black 
lines in each panel. Bottom: the quantity $d\ln P/d\ln r$ at 
t = 8000 and 10000 orbits in each case. 
The violet dashed line in the top panel denotes 
$\Sigma_{g}$ at t = 10000 orbits in the gas case. The dashed light blue 
line in the top and bottom panels denotes the initial value of 
the presented quantity while in the middle panel it indicates 
$\Sigma_{d-e}$. In the bottom panel the orange dashed horizontal line shows the
locations of the zero pressure gradient in the disk. 
The red and black dashed vertical lines in each panel separately indicate the 
positions of the inner and outer planets at t = 8000 and 10000 orbits.}
\label{fig:two-dsden-com-v5-1}
\end{figure*} 

\begin{figure*}[htbp]
\centering
\includegraphics[width=1.0\linewidth]{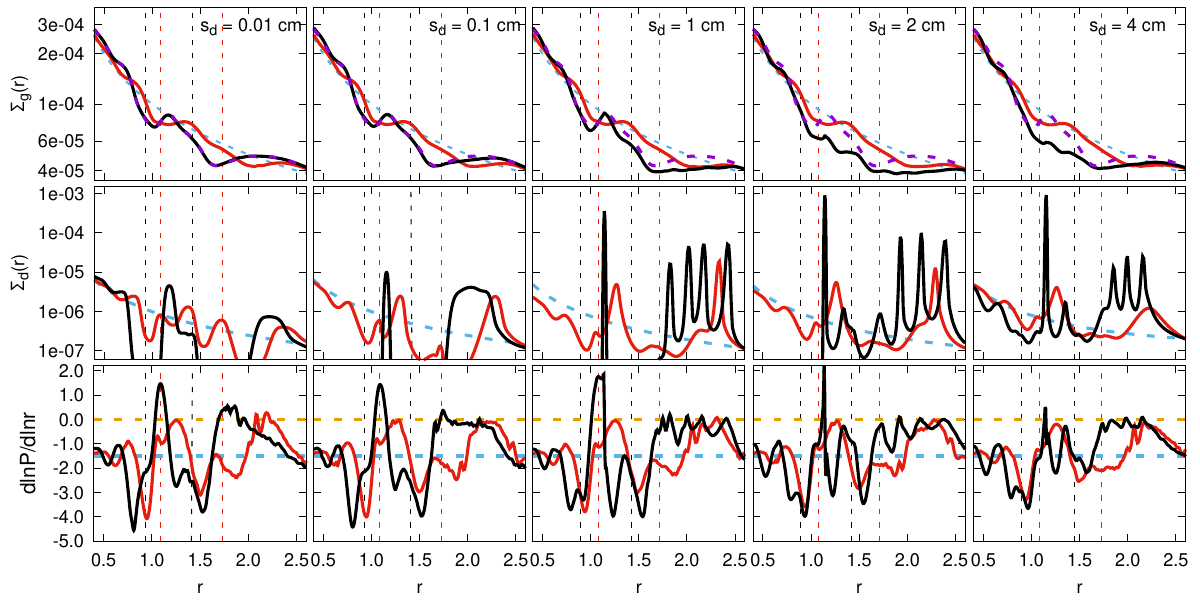}
\caption{The same as Figure \ref{fig:two-dsden-com-v5-1} 
at t = 8000 (red line)  and 18000 (black line) orbits. 
The violet dashed line in the top panel denotes 
$\Sigma_{g}$ at t = 18000 orbits in the gas case. 
}
\label{fig:two-dsden-com-v5-2}
\end{figure*}

The gas surface density $\Sigma_{g}$ as a function of $r$ in 
the disk close to the planet locations at t = 8000 (red line) and 10000 
orbits (black line) in each of
the simulation with a given grain size is presented 
in the top panel of Figure~\ref{fig:two-dsden-com-v5-1}. 
The dashed light blue line denotes the initial value of the gas surface density
and the violet dashed line indicates the
gas surface density at t = 10000 orbits in the disks without dust grains.

At t = 8000 orbits, the planets are in the 2:1 MMR. Both planets
opened shallow partial gaps in the disks in which they 
are embedded. 
The gas surface density profiles of the disks with the grains of different
size 
are similar to each other and to the surface density profile of the purely
gaseous disk.
After 2000 orbits of further evolution, at t=10000 orbits, the situation
is not much different, there
are only small differences in the depths of the gaps and
the amount of gas present in the co-orbital regions of the
planets.

Comparing those surface density profiles with the results for the 
corresponding isolated 
planet cases, we checked that the depth 
of the partial gaps of the inner planet, at the same stage of the evolution,
are deeper in the two-planet 
simulations, while for the outer planet, the partial gap depths in
the two-planet and the corresponding isolated planet cases are   
similar to each other.

The noticeable distinctions in the gas surface density profiles appear
only after 10000 orbits of the evolution.
At 18000 orbits, the planets in the gas case, 
and in the cases $s_{d} = 0.01$ and 0.1 cm passed already through 
the 2:1 MMR, those in the $s_{d} $ = 1 cm case 
are still locked in the resonance and those
in the $s_{d}$ = 2 cm and 4 cm cases undergo divergent migration. The
surface density profiles resemble in shape very wide double-gaps
or single shallow gaps as can be seen in the top panel of 
Figure~\ref{fig:two-dsden-com-v5-2}. The gaps are 
visibly deeper than those present in the disks at 8000 orbits. 
The gas distributions in the disks without dust and 
with dust grains with $s_d$ = 0.01 and 0.1 cm remain 
similar to each other after the same number of orbits for the whole duration
of the calculations. 
In the $s_{d}$ = 1 cm case, the gas distribution around 
the partial gap of the inner planet remain similar to that of the gas case 
but the partial gap of the outer planet after 10000 orbits becomes wider and 
deeper.
Moreover, the partial gaps in the disks with 
larger dust grains ($s_{d} = 2$ cm and $s_{d} = 4$ cm cases) at t = 18000
orbits differ significantly from 
the gas distributions in the disks with the smaller grains taken 
at the same time. 
In particular, the gas in the region between
the planets is depleted in the disks with large dust grains, so the two gaps
merge forming one wide shallow gap. 

In summary, the 
presence of small dust particles (less or equal 0.1 cm in size) 
in the disk 
does not change much the density profile relative to the 
gas surface density in the purely gaseous disk. 
Instead, when $s_{d}$ is larger, the gas evolution can be 
significantly affected by the dust dynamics and thus the profile of 
$\Sigma_{g}$ does not resemble that of the purely gaseous disk.  
To illustrate our conclusion, we indicate the surface density profile
of the purely gaseous disk at t = 18000 orbits in the top panel of 
Figure~\ref{fig:two-dsden-com-v5-2} (violet dashed line). 

\subsubsection{Evolution of the dust surface density}

After showing the way in which the dust grains 
affect the gas surface density of the disks in our calculations,
we move on to the description of the dust density distribution  
and the relation between gas and dust evolution. 
For this purpose, we draw also the azimuthally averaged dust surface 
density at t = 8000 (red line) and 10000 (black line) orbits in the 
vicinity of two planets in the simulations with various $s_{d}$ 
(middle panel of Figure~\ref{fig:two-dsden-com-v5-1}) and 18000 (black line)
orbits (middle panel of Figure~\ref{fig:two-dsden-com-v5-2}). 
It can be noticed that at t=8000 orbits, 
both planets carved the double-shaped gaps (inner and outer planets when 
the grains have $s_d=0.01$ or 0.1 cm and only inner planet when the 
grains are larger)
or the wide and shallow gaps (only outer planet when the grains have 
$s_d$ = 1, 2 or 4 cm) in the disk along their orbits. 
Apart from $s_{d} =$ 0.01 cm case, the gaps formed by 
the outer planets in the disks with the smaller grains are deeper while 
the dust particles are accumulated on the gap edges.
After additional 2000 orbits not many changes can be seen in the disk
with the smallest grains, but already for the disk with $s_d=0.1$ cm one
can notice that the gap around inner planet is very deep and that more
dust accumulates between planets. 
For the disks with larger grains the gaps around the inner planets become
deeper, especially for the inner planet in the $s_{d}$ = 1 cm 
case, and the rings of dust between planets contain more dust. 
The gaps around outer planets do not change much, but the amount of dust
at the outer edge of the gaps increases. 
At 18000 orbits the gaps are getting deeper
(except for the gaps around outer planets in $s_{d}=2$ and 4 cm cases), 
the dust concentration peaks
located between planets are higher and the dust features exterior to 
outer planet are wider and contain more dust.

It is of interest to discuss the surface densities profiles in the
vicinity of planets migrating in the disk in terms of the dust traps formed 
in the region where a pressure maximum is generated 
or the inflection points are present, see for example 
\cite{2024MNRAS.528.6538P}. To this aim, in the bottom panels of 
Figure~\ref{fig:two-dsden-com-v5-1} and Figure~\ref{fig:two-dsden-com-v5-2}
we show the logarithmic gradient of the azimuthally averaged gas pressure
$d\ln P/d\ln r$ as a function of $r$ at t = 8000, 10000 and 18000 orbits.
The positions of the inner and outer planets at each moment of time 
are indicated by the red and black dashed vertical lines in each panel,
respectively.

At t = 8000 orbits, the pressure gradient in all five cases has two inflection
points and the corresponding rings of dust can be seen close to their location,
confirming that the inflection points acts as the dust traps.  
As the evolution goes on, at t = 10000 orbits in the disk with $s_d$ = 0.01 and
0.1 cm the pressure gradient becomes positive 
in the region between two planets and at t=18000 orbits it is positive in all
five cases. 
In the disk with the small dust grains the dust surface density in the rings is lower than 
the gas surface density and for the large grains it is reaching the same
value as the gas surface density at about t=10000 orbits ($s_{d} = 2$ and 4 cm) or 
t = 11000 orbits ($s_{d} =$ 1 cm) and later on becomes higher than that.

The accumulation of dust located exterior to the outer planet,
centered around $r=2.0$, corresponds
to the region where the pressure gradient is close or equal to zero 
(the pressure there is approximately constant).
In the exterior region of the outer planets in the $s_d$ = 1, 2 and 4 cm
cases several dust rings are formed, instead of a flat structure as 
in the $s_{d} = 0.1$ cm or 0.01 cm cases. 
It can be noticed that in the disks with large grains, as time proceeds, 
more dust rings are formed outside the outer planet orbits 
and their peak values can exceed $\Sigma_{g}$.
As we have already mentioned, whenever the dust-to-gas ratio is bigger
than 1 the treatment of dust, adopted here, is not fully justified.

The dust distribution around two super-Earths 
in a low viscous disk has been studied in \cite{2025AA...703A.270R}. In their 
Model 4 considering an inner planet with $10 M_{\oplus}$ and an outer planet 
with $5 M_{\oplus}$, several narrow dust rings are formed at the edges of the gaps 
due to the gas pressure bump in the disk with $s_{d}$ = 200 $\mu$m, 1.6 mm and 2.6 cm
while the density at the peak of dust ring between two planets is higher for the larger $s_{d}$. 
The dust rings observed in our cases have similar properties to their results. 

\subsection{The torques from the disk material acting on the planets} 
\label{subsec:torque}

In this part, we investigate how the torques from the gas and dust acting 
on the planets evolve in time to explore further the question about
the effects of dust dynamics on the planet migration and to find out the
reason for which the divergent migration occurs in the disks with the large 
dust grains, namely $s_d$ = 2 cm and 4 cm.

\subsubsection{The gas torque}

\begin{figure}[htbp]
\centering
\includegraphics[width=1.0\linewidth]{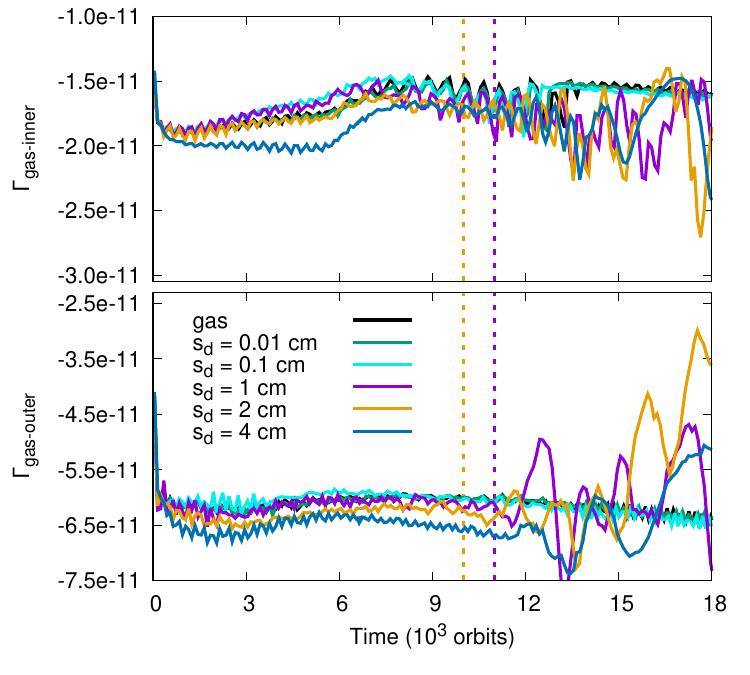}
\caption{The gas torque acting on the inner and outer planets as a function 
of time in the cases with various $s_{d}$ and in the gas case.
The orange and violet dashed vertical lines have the same meaning as in 
Figure~\ref{fig:evo-two-v5}. }
\label{fig:two-gas-torque-v5}
\end{figure}

In Figure~\ref{fig:two-gas-torque-v5}, we show the evolution in time of
the gas torques acting
on the planets embedded in the disks with different dust content.
The torque, $\Gamma_{\rm gas-inner}$, 
acting on the inner planet 
immersed in the disk with the dust grains $s_{d} = 0.01$ cm in size 
(green line in the top panel) evolves, 
in a similar manner as that acting in the purely gaseous 
disk (black line in the top panel) 
during the whole time of the calculations.
The same is true for the evolution
of the torques acting on the outer planet $\Gamma_{\rm gas-outer}$ (green
and black lines in the bottom panel). 
The torque in the disks with the $s_{d} = 0.1$ cm grains (cyan line) 
is also not much different than those two, described above, 
but only
starting from t = 8000 orbits, when the libration overstability is 
fully developed.
The two vertical dashed lines are not relevant for the
disks with the small dust grains, because the dust-to-gas ratio, $\epsilon $,
in such
disks is always and everywhere much less than 1 for the duration of 
our simulations. 

The effects of dust dynamics on the gas torque are
more pronounced when $s_{d}=$ 1 cm (violet lines). 
At the beginning $\Gamma_{\rm gas-inner}$ follows closely the evolution of
the torque acting on the planet in the case of the disk with the 0.1 cm grains,
but then at about 7500 orbits 
its absolute value, $|\Gamma_{\rm gas-inner}|$, becomes larger than 
the equivalent value of the torque in the gas 
case and also in the disks with the 0.01 and 0.1 cm grains. 
At the time of crossing the violet vertical dashed line when $\epsilon $ 
reaches the value of 1 in at least one of the dust rings formed in the disk during the evolution,
the oscillations of the $|\Gamma_{\rm gas-inner}|$ become larger and continue
to grow till the end of the calculations. 
Instead, $|\Gamma_{\rm gas-outer}|$ is similar to the absolute value of the
gas torque in 
the gas case till t = 11000 orbits (when $\epsilon = 1 $) and then the 
amplitude of its oscillations
becomes much larger, but 
its averaged value is comparable to the value of the gas torque in 
the gas case.

In the $s_{d}$ = 2 cm case (orange lines),
$|\Gamma_{\rm gas-inner}|$ is noticeable larger than the equivalent
absolute value of the torque in the gas case after 8000 orbits 
and follows approximately the evolution of the torque in the case of
1 cm grains.
In the case of outer planet, $|\Gamma_{\rm gas-outer}|$  
is larger than the absolute value of the torque acting in the disks with 
the grains smaller than 2 cm already from the beginning of the calculations.
Starting from t = 10000 orbits ($\epsilon =1$ for 2 cm grains), the gas 
torques acting on the inner and outer planets begin to oscillate with large 
amplitudes. 
The averaged value of $|\Gamma_{\rm gas-inner}|$ 
increases while $|\Gamma_{\rm gas-outer}|$ largely decreases. This results in 
faster migration of the inner planet and slower migration of the outer planet. 
The final outcome is the divergent migration.
In the $s_{d} = 4$ cm case (dark blue lines), both $|\Gamma_{\rm gas-inner}|$ 
and $|\Gamma_{\rm gas-outer}|$ are larger than the equivalent 
values in the gas case 
during most of the calculation time. At $t \sim 12000$ orbits, the 
amplitude of the oscillations of two gas torques begin to increase. 
$|\Gamma_{\rm gas-inner}|$ increases and $|\Gamma_{\rm gas-outer}|$ decreases, 
which causes the planets migrate divergently as in the $s_{d} = 2$ cm case.

\subsubsection{The dust torque}

\begin{figure*}[htbp]
\centering
\includegraphics[width=1.0\linewidth]{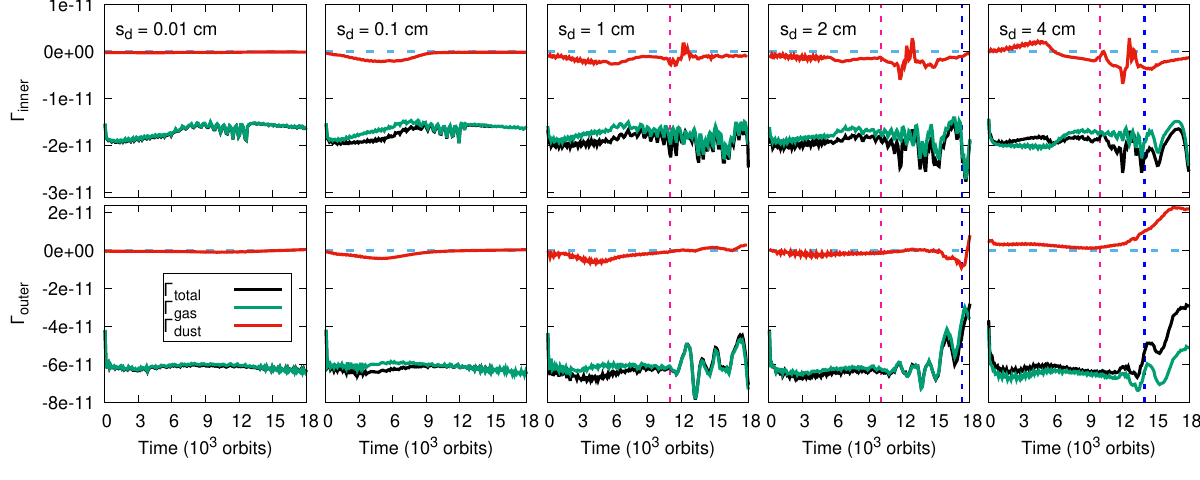}
\caption{The total torque, the gas torque and the dust torque acting on the
inner (top panel) and outer (bottom panel) planets as a function of time
in the simulations with
different $s_{d}$. The pink dashed vertical lines in
the panels for $s_d$ = 1, 2 and 4 cm indicate 
the moments when $\epsilon$ at the peak of the dust ring between planets exceeds unity.
The blue dashed vertical lines represent the moments when the occurrence of the divergent migration takes place.
}
\label{fig:two-total-torque-v5}
\end{figure*}

To illustrate the contribution of the dust present in the disks 
to the total torque acting
on the inner and outer planets, we draw the evolution in time 
of the total torque $\Gamma_{\rm total}$, 
the gas torque $\Gamma_{\rm gas}$ and the dust torque $\Gamma_{\rm dust}$ 
in Figure~\ref{fig:two-total-torque-v5}.

In the $s_{d} = 0.01$ cm case, 
the dust torque acting on each of the planets is nearly zero, 
which leads to $\Gamma_{\rm total} \sim \Gamma_{\rm gas}$ during the whole 
calculation.

In the $s_{d} = 0.1$ cm case, the dust torque acting on each planet 
becomes more negative with increasing time till the moment 
when the planets enter the 2:1 resonance ($t \sim 5300$ orbits).  
During the resonance this trend is reversed and the dust torque 
decreases in its absolute value approaching zero at  
$t \sim 10000$ orbits. Then it remains close to zero till the end of the 
calculations. Interestingly, towards the end
of the calculations the dust torque acting on the outer planet, being
always very close to zero, 
becomes slightly positive. 
In consequence $\Gamma_{\rm total} \sim \Gamma_{\rm gas}$ till the end 
of the calculation and we can conclude that in these two cases ($s_{d} = 0.01$ 
and 0.1 cm), considered in this work,
the dust dynamics does not have substantial 
influence on the planet migration.

In the $s_{d}$ = 1 cm case, the dust  
torque acting on both planets is similar to that in the $s_{d}$ = 0.1 cm case
till t = 8000 orbits. 
Later on, at t = 11000 orbits, when the dust-to-gas ratio, $\epsilon$, 
in the dust ring growing between two planets 
reaches unity (the vertical dashed pink line),
$\Gamma_{\rm dust}$ for the inner planet exhibits oscillations, 
then becomes positive for a little while and finally changes back to be negative,
remaining always close to zero.
Meanwhile, $\Gamma_{\rm dust}$ for the outer planet oscillates around zero and
towards the end of the calculations 
becomes positive, but its value is not far from being zero.  
Therefore, the total torque acting on two planets are dominated by their 
gas torques. 
The difference between the $s_{d} = 0.1$ cm and $s_{d} = 1$ cm cases 
in the gas torque behavior has been already discussed
and shown in detail in Figure~\ref{fig:two-gas-torque-v5}.

In the $s_{d} = 2$ cm case,  
$\Gamma_{\rm dust}$ for both planets, before the occurrence of the resonance,
is less negative than for the $s_{d} = 0.1$ cm and 1 cm cases. 
However, later on the gas torque evolution proceeds in a similar qualitative way
as for the smaller grains till the time when 
$\epsilon$ exceeds unity.  
After that time the dust torque evolution 
for both planets are similar to the $s_{d}$ = 1 cm case except that 
$\Gamma_{\rm dust}$ for the outer planet decreases 
to be negative since t = 15000 orbits and increases to be positive 
again at the end of the simulations.
$\Gamma_{\rm total} \sim \Gamma_{\rm gas}$
till t = 17500 orbits (the blue vertical dashed line) for both planets and for 
nearly the entire time of the calculations. 
After that, $\Gamma_{\rm dust}$ for the inner planet becomes slightly 
more negative but still close to zero. For that reason, the dust torque
gives a very small contribution to $\Gamma_{\rm total}$.
Instead, $\Gamma_{\rm dust}$ for the outer planet grows fast and
becomes positive giving
a significant contribution to  $\Gamma_{\rm total}$. In consequence
the outer planet 
migrates inward slower and the divergent migration occurs. 

In the $s_{d} = 4$ cm case,  
$\Gamma_{\rm dust}$ 
is positive for both planets before entering the 2:1 resonance and 
even a little bit further till $t \sim 6000$ orbits. 
This results in  
$|\Gamma_{\rm total}|$ for two planets being smaller than their 
$|\Gamma_{\rm gas}|$ and thus the migration rates of the planets are 
slower than in other cases in the same moment of time. After the planets 
enter into the 2:1 MMR, $\Gamma_{\rm dust}$ for the inner planet becomes 
negative while $\Gamma_{\rm gas}$ increases. $\Gamma_{\rm total}$ for the
inner planet is decreasing as a result of the negative dust torque, which 
means that the dust particles push the inner planet to migrate inward faster.
At the time when 
the peak of the dust ring between two planets has $\epsilon$ bigger than unity,
$\Gamma_{\rm dust}$ for the inner planet begins to oscillate 
significantly. That is similar to what we have observed in the  $s_{d} = 2$ cm case. 
Instead, $\Gamma_{\rm dust}$ for the outer planet increases rapidly, which 
leads to the increase of $\Gamma_{\rm total}$ acting on the outer planet. 
Therefore, 
the inward migration of the outer planet becomes slower 
as in the $s_{d} = 2$ cm case,
and the planets migrate divergently.
 
\subsubsection{What is the reason for the divergent migration?}

The torque evolution shown in Figure~\ref{fig:two-total-torque-v5} indicates 
that the divergent migration in the $s_{d} = 2$ cm and $s_{d} = 4 $ cm 
cases should be related to the formation of the dust ring between two planets.
In the case of $s_{d} = 2 $ cm, when $\epsilon > 1$ 
at the maximum of the dust surface density in the ring, 
then $\Gamma_{\rm gas}$ acting on the 
outer planet increases and in the case of $s_{d} = 4$ cm, 
it is $\Gamma_{\rm dust}$ on that planet which increases. 
In consequence, the total torque acting on the outer planet, 
$\Gamma_{\rm total}$, increases and the inward migration of the 
outer planet slows down.

\begin{figure}[htbp]
\centering
\includegraphics[width=1.0\linewidth]{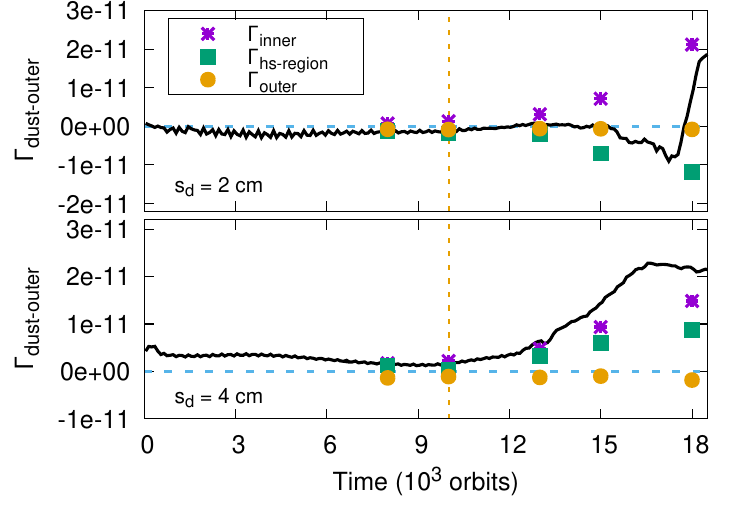}
\caption{The dust torques from different parts of the disk acting on the
outer planet in the simulations with $s_{d} = 2$ cm (top panel)
and 4 cm (bottom panel). See text for more details.
The orange dashed line has the same meaning as in
Figure~\ref{fig:evo-two-v5}.
}
\label{fig:two-dust-torque-outer-sep}
\end{figure}

To demonstrate that 
the dust ring between the planets is responsible for slowing down 
the outer planet migration, we calculate the torque generated from the dust particles 
in the interior region of the disk with $r < r_{p2} - x_{s}$, the horseshoe 
region with $r_{p2} - x_{s} \leq r \leq r_{p2} + x_{s}$ and the exterior 
region with $r > r_{p2} + x_{s}$ acting on the outer planet in these two 
cases, where $r_{p2}$ is the position of the outer planet, 
and $x_{s}$ is 
the half-width of the horseshoe region of the outer planet calculated 
as $x_{s} = 1.2r_{p2}\sqrt{q_{2}/h}$. 
The dust torque generated from this dust ring is included in the 
torque of the interior region.

In Figure~\ref{fig:two-dust-torque-outer-sep} we draw the dust torque 
from each region at several moments of time and the total dust torque 
acting on the outer planet. In the top panel for the $s_{d} = 2 $ cm 
case, we can see that since t = 10000 orbits, $\Gamma_{\rm inner}$ 
increases significantly. This corresponds to the increase in the peak value 
of $\Sigma_{d}$ of the dust ring located in this region. 
At the same time, $\Gamma_{\rm outer}$ does not change significantly, which
means that the several dust rings present in the exterior region with
$\epsilon > 1$ do not contribute to the total dust torque as much as the 
dust ring between the planets.
Similar behavior can be seen also in the bottom panel for the $s_{d} = 4 $ cm 
case indicating that also in that case $\Gamma_{\rm inner}$ increases 
due to the formation of the dust ring between the planets 
and $\Gamma_{\rm outer}$ stays around zero.

\begin{figure}[htbp]
\centering
\includegraphics[width=1.0\linewidth]{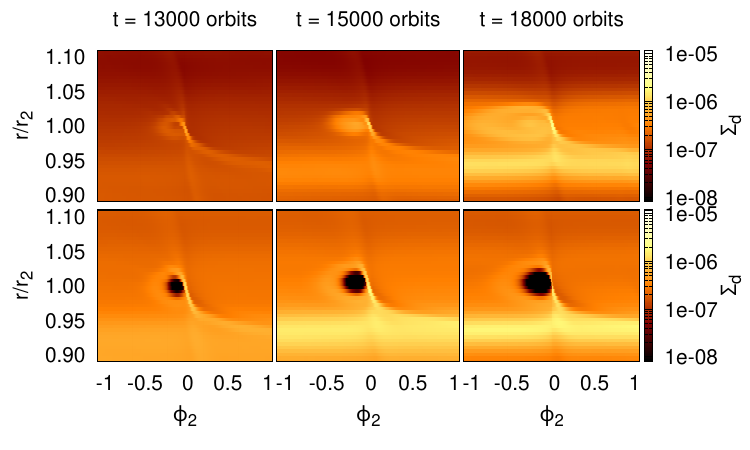}
\caption{The contour plots of $\Sigma_{d}$ in the vicinity of the outer
planet at t = 13000, 15000 and 18000 orbits in the $s_{d} = 2$ cm case
(top panels) and the $s_{d} = 4$ cm case (bottom panels).
}
\label{fig:two-outer-dust-contour-v5}
\end{figure}

Furthermore, it is noteworthy that
$\Gamma_{\rm hs-region}$ decreases 
in the $s_{d} = 2 $ cm case but increases in the $s_{d} = 4$ cm case. 
To explore why $\Gamma_{\rm hs-region}$ evolves differently, 
we draw the contour plots of $\Sigma_{d}$ in the vicinity of the outer 
planet at t = 13000, 15000 and 18000 orbits in these two cases 
in Figure~\ref{fig:two-outer-dust-contour-v5}. 

We find that in the $s_{d} = 2$ cm case
(top panels), the dust 
void on the left side of the planet at t = 13000 
orbits is small, much smaller compared to its size at the earlier
times of the evolution (not shown in the figure). As the calculations
proceed the dust particles fill in the void and the $\Sigma_{d}$ in this
region becomes higher than the surroundings. 
At t = 18000 orbits, the size of this dust structure becomes 
larger with even higher $\Sigma_{d}$. 
Such asymmetric dust distribution in the co-orbital region of 
the outer planet generates the negative torque, which 
makes $\Gamma_{\rm hs-region}$ to decrease in the $s_{d} = 2$ cm case.

In the bottom panel for the $s_{d} = 4 $ cm case, we notice that the dust 
void on the left side of the outer planet is present at t = 13000 
orbits. Moreover, the size of the dust void increases along with 
the evolution time and thus the dust in the co-orbital region 
generates the positive torque \citep{2018ApJ...855L..28B}. 
Therefore, $\Gamma_{\rm hs-region}$ increases in the $s_{d} = 4$ cm case.

Based on this analysis, we sum up that the dust torque 
generated from the dust ring present between the planets contributes
significantly  
to the total torque of the outer planet in the $s_{d}$ = 2 cm and 4 cm cases, 
when $\epsilon$ at the dust peak exceed unity. 
This helps to slow down the inward migration of the outer planet. 
Moreover, this dust ring affects 
also the inner planet migration, because
$\Gamma_{\rm dust}$ acting on the inner planet is negative during most of 
the time of the calculations. Therefore, the inner planets migrate faster 
if the substantial amount of dust is accumulated between planets.
This plays a role in the occurrence of the divergent migration
(see the top-middle panel of Figure~\ref{fig:evo-two-v5}).
Moreover, the co-orbital asymmetric substructures as the dust voids  
can have non negligible contribution to the total torque acting
on the planet. 
Therefore, the coupled effects of gas and dust  can efficiently modify
the planets migration.  
 
\section{Orbital evolution of two super-Earths near the 2:1 resonance in 
the viscous disk of gas and dust}
\label{sec:two-planets-v3}

\begin{figure*}[htbp]
\centering
\includegraphics[width=1.0\linewidth]{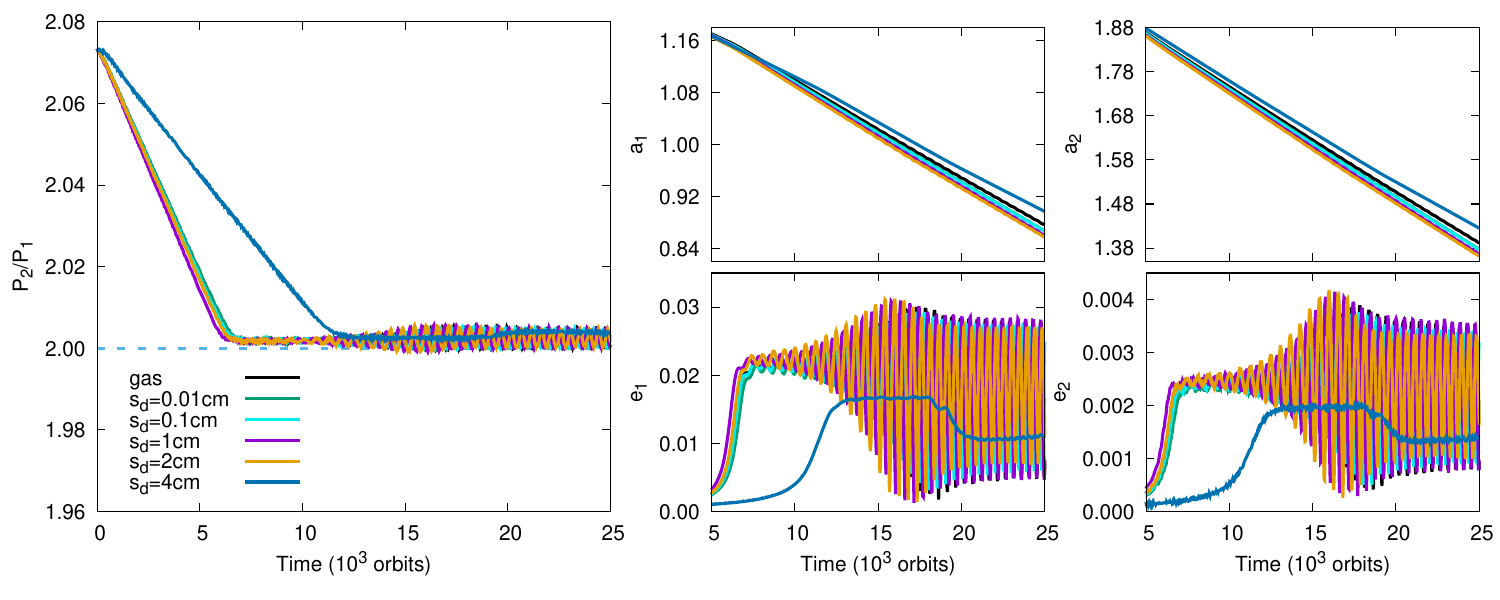}
\caption{Evolution of the period ratios ($P_2/P_1$), the semi-major axes 
($a_1$, $a_2$) and eccentricities ($e_1$, $e_2$)
of two planets in the purely gaseous 
disk and in the disks
of gas and dust with different $s_{d}$. For the semi-major axes and 
eccentricities 
we show the most relevant part of the evolution, omitting the first 5000 orbits
during which no significant differences can be seen.
The viscous parameter $\alpha_{0}$
is taken to be $10^{-3}$. The horizontal dashed light blue line 
indicates the position of the 2:1 MMR.}
\label{fig:evo-two-v3}
\end{figure*}

The results of our simulations with $\alpha_{0} = 10^{-5}$ show that 
the convergent migration of two super-Earths embedded in a low-viscosity 
disk of gas and dust leads to a resonance capture into 
the 2:1 commensurability. The resonance locking is temporary if
planets evolve in the disks with dust grains having $s_d \le 0.1$  
cm. The planets leave the commensurability and continue their convergent 
migration. If the disk contains 1 cm dust grains then the planets
stay in the resonance till the end of the
simulations and in the disks with the 2 and 4 cm dust grains the planets
leave the commensurability and undergo the divergent 
migration (Figure~\ref{fig:evo-two-v5}).
To investigate how this picture will change after adopting significantly 
higher viscosity, we repeat the simulations presented in the previous section
with $\alpha_{0} = 10^{-3}$. The expectation is that it will be more 
difficult for the super-Earths 
to carve the partial gaps in the disk and the dust diffusion mechanism 
will be stronger.

The results of our simulations with $\alpha_{0} = 10^{-3}$ are shown 
in Figure~\ref{fig:evo-two-v3}. For the purely gaseous disk they are in
a good agreement with those given in  
\cite{2024AA...686A.277A} for the same value of viscosity.   
In all cases (purely gaseous disk, disks with the dust grains with 
$s_d=0.01$, 0.1, 1, 2 and 4 cm) 
both planets migrate inward till the end of the calculations. 
The migration rate of each planet in the $s_{d} = 4$ cm case is 
the slowest among
all others. The second slowest is the migration in the purely gas case and then
the migration rates increase with increasing size of the dust present in the
disk. However, the difference between the evolving semi-major axes in the  
gas and $s_d=0.01$ cm cases is so small that in Figure~\ref{fig:evo-two-v3} 
they practically overlap each other.
At the beginning of their evolution, planets migrate convergently 
and  enter into the 2:1 MMR. 
In the purely gas case and the cases with $s_{d}$ = 0.01, 0.1, 1 and 2 cm 
the capture takes place at $t \sim 6000$ orbits and  
both eccentricities are excited as seen in the middle and right 
bottom panels of 
Figure~\ref{fig:evo-two-v3}. 
After that, the planets stay in the resonance till the end of the calculations,
exhibiting after 20000 orbits (apart from the case $s_d$ = 4 cm) 
a limit cycle behavior. The limit cycle manifests itself by the oscillations
of the period ratios, eccentricities and resonance angles around an 
equilibrium value with constant amplitude. Such behavior is clearly seen
in the evolution of eccentricities in Figure~\ref{fig:evo-two-v3}.
Comparing the outcome of the simulations performed with 
$\alpha_{0} = 10^{-3}$ and those with $\alpha_{0} = 10^{-5}$ presented 
in the previous section, it can be
noticed that the relative migration rates between planets are slower in the 
disks with higher viscosity. This is particularly evident in the case of the
disk with $s_{d} = 4$ cm, where planets arrive to the 2:1 MMR resonance only  
after about 11000 orbits. The eccentricities 
$e_{1}$ and $e_{2}$ are respectively excited to 0.015 and 0.0019. They are 
somewhat lower than in the cases with the smaller dust grains. 
Then, at about $t \sim 19000$ orbits the period ratio slightly increases
and the eccentricities decrease to $0.01$ and $0.0012$, respectively.
These values are maintained till the end of the calculations and the
planets remain in the 2:1 MMR. 

\begin{figure}[htbp]
\centering
\includegraphics[width=0.9\linewidth]{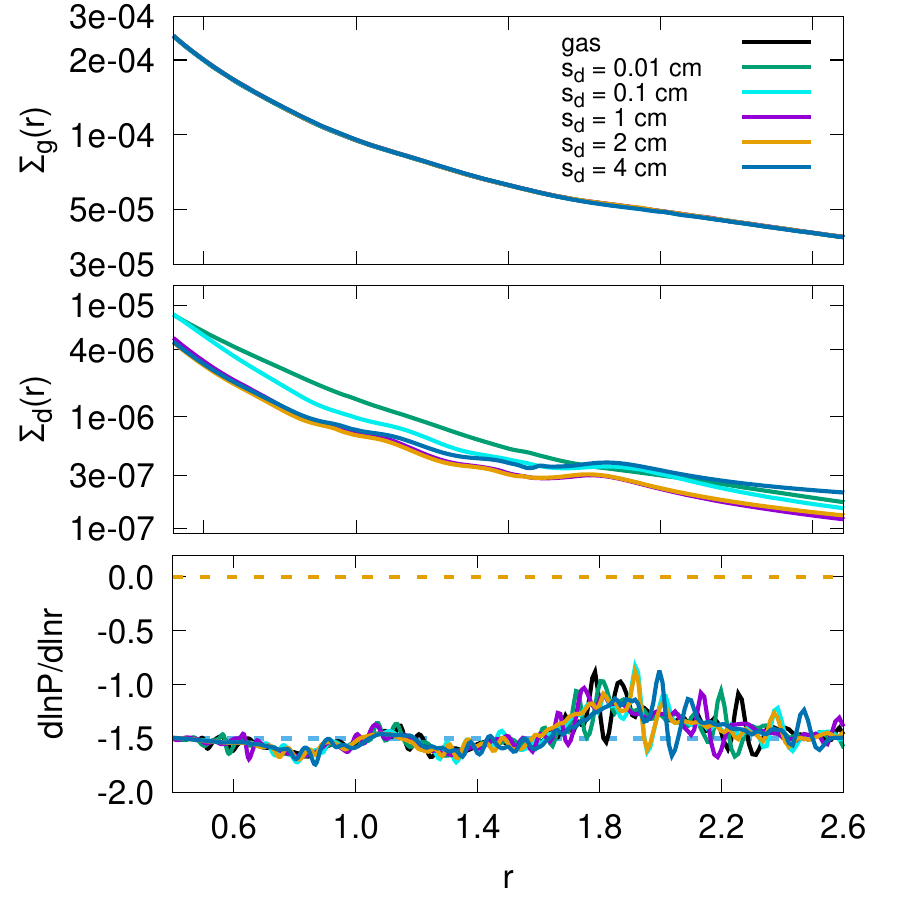}
\caption{Top and middle: The azimuthally averaged gas and dust surface density of the 
disk in the simulations of the gas case and the cases with 
various $s_{d}$ at t = 18000 orbits. 
Bottom: the quantity $d\ln P/d\ln r$ at t = 18000 orbits.
The dashed light blue horizontal line denotes the initial value of the logarithmic 
pressure gradient and the orange dashed line shows where this quantity is equal zero. 
The viscous parameter is taken to be $\alpha_{0} = 10^{-3}$. }
\label{fig:two-sden-dsden-v3}
\end{figure} 

In order to make a comparison between the disk surface density 
profiles shaped by the planet pair evolution in the disks
with two different viscosities, namely $\alpha_0 = 10^{-5}$ presented in 
Figure~\ref{fig:two-dsden-com-v5-2} and $10^{-3}$ given in
Figure~\ref{fig:evo-two-v3}, we show
in Figure~\ref{fig:two-sden-dsden-v3}, the azimuthally averaged gas surface 
density $\Sigma_g$ (upper panel), the azimuthally averaged dust surface 
density (middle panel)
and the logarithmic pressure gradient (bottom panel) 
in the vicinity of two planets at t = 18000 orbits.
The comparison shows that the gas distribution around two planets 
in a viscous disk with various $s_{d}$ are all quite similar to the gas case.
Moreover, as expected, the planets open only very shallow partial gaps (not
visible in the scale adopted in the figure, but clearly seen in the gas
surface distribution if normalized by the initial gas density profile). 
Two inflection points present in the gradient of pressure distribution
(bottom panel)
result in formation of two small bumps in the dust density distribution (middle
panel). They do not grow in time, as in the case of the  dust 
structures in the disk with $\alpha_0 =10^{-5}$ and
$\Sigma_{d}$ is always less than $\Sigma_{g}$ in the whole disk region. 

\section{Discussion}
\label{sec:discussion}

\subsection{Properties of the dust rings formed by two super-Earths 
in a low-viscosity disk}

Our simulations show that the migration of two super-Earths in 
a low-viscosity disk ($\alpha_{0} = 10^{-5}$) can lead to the formation
of one or more dust rings in the vicinity of the planetary orbits.
The number of the dust rings and their properties such as 
their locations, widths and maximum values of the dust surface density depend
on the dust particle size $s_{d}$. 

\begin{figure*}[htbp]
\centering
\includegraphics[width=1.0\linewidth]{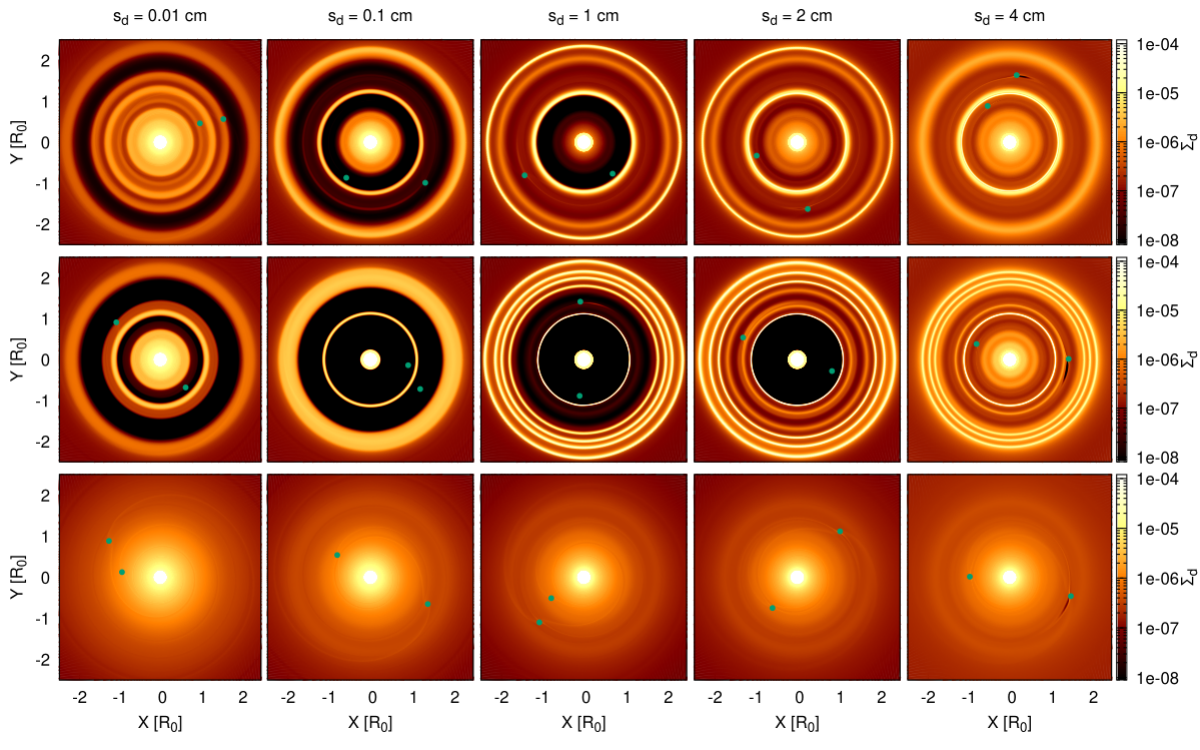}
\caption{The contour plots of dust surface density $\Sigma_{d}$  
in the simulations with various $s_{d}$ and two values
of the viscosity parameter $\alpha_0$: $10^{-5}$ (at 10000 orbits - upper 
panel and at 18000 orbits - middle panel) and $10^{-3}$ (at 18000 orbits -
lower panel). 
From left to right $s_{d}$ is taken to be 0.01, 0.1, 1, 2 
and 4 cm, 
respectively. The green circles denote the positions of two planets.}
\label{fig:two-contour-disk-v5}
\end{figure*}

In Figure~\ref{fig:two-contour-disk-v5}, we present the contour plots 
of $\Sigma_{d}$ in the nearly inviscid disks at two moments of time, namely
t = 10000 orbits (upper panel) and t = 18000 orbits (middle panel) and in the 
viscous  disks at t = 18000 orbits
(lower panel) with
the dust grains of different size $s_d$ (from left to right: $s_{d}$ =
 0.01, 0.1, 1, 2 and 4 cm). 
The two planet locations are marked with the green circles. 
The super-Earths migrating in the nearly inviscid disks create 
a number of dust structures: rings, gaps, central cavities 
and non-axisymmetric features.

In the $s_{d} = 0.01$ cm case, the mild not very narrow dust ring 
is present in the low-viscosity disk at the location just outside the 
initial position of the outer planet ($r=2.0$) at t = 10000 orbits. 
Each planet carved two double gaps on both side of its orbit.
The gap exterior to the outer planet orbit is the deepest of all. The 
super-Earths are located at the local not very pronounced dust rings.
The dust surface density in the innermost region
of the disk is similar to its initial value, no central cavity is formed. 
After 8000 orbits of the evolution, the outermost ring becomes wider
without changing significantly its position. The outermost gap gets also
wider and the outer planet is located very close to the inner 
edge of the gap.
The inner planet orbits in the deep gap and the noticeable dust ring is
formed between the planets. 
Also at that time the central cavity is not present.
In the viscous disk we can see just spiral disturbances caused by the
migrating super-Earths embedded in it.

When the low-viscosity disk contains larger grains ($s_d=0.1$ cm) then already
at t = 10000 orbits two deep gaps are formed. The outermost ring has larger
surface density than the one in the case of smaller grains. The planets are 
located in the gaps and there is an asymmetric filamentary dust structure 
started at the 
outer planet position. 
The two gaps are divided from each other by  
the dust ring, which is more narrow in comparison with the previous case.
At t = 18000 orbits, the outermost ring gets much wider and the ring between 
planets becomes even more narrow. 
Moreover, the dust from the innermost region of the disk is depleted
and the gap, where the inner planet is located, extends practically till the
inner edge of the disk, forming the central cavity. In addition, both gaps 
are clean: there are no 
noticeable  asymmetric dust structures created by the outer planet. In the 
case of the viscous disk, a very shallow gap is formed, in which the 
outer planet is located. The only 
other dust structures present in the dust surface distribution
are the spiral arms excited by planets. 

In the case of larger dust grains present in the gaseous 
low-viscosity disk ($s_{d}$ = 1 cm and 2 cm), instead of the outermost wide 
ring seen in the disks with smaller grains at t = 10000 orbits, we observe 
two narrower rings. The one on the outside has the noticeably bigger 
dust surface density than the other. The planets orbit each in its own gap. 
The partition between those two gaps is a narrow ring formed between the 
planets.  
In the disk with 1 cm grains, the inner gap is deep while the outer gap 
is shallow and some irregular dusty material is present inside it. 
In the innermost part of the disk the dust is mostly removed but the central 
cavity formation is not completed yet. 
In the disk with 2 cm grains, both gaps are shallow filled in with the dusty
material. In the innermost part of the disk the dust
is partially depleted but the process of the central cavity formation is still
in its early phase.

After further 8000 orbits of the evolution of the system, the outer gap in the
disk with 1 cm grains becomes deep. However, still the presence of a dusty 
filamentary structure starting at the planet position can be noticed. 
Outside this gap four dust rings are present. 
At the location of the three most exterior ones, the maximum of the dust 
surface density is slightly bigger than the gas surface density ($\epsilon >1$)
and for the fourth one $\epsilon <1$, even if not by much.
The narrow dust ring formed between the orbits of the planets is at this time 
even thinner with $\Sigma_{d}$ far exceeding $\Sigma_{g}$. 
In the innermost part of the disk the central cavity is fully formed.
The central cavity is also present in the disk with 2 cm grains, as well as
the very narrow dust ring between two planets with $\epsilon $ exceeding unity.
The difference is, though, that just outside this ring there is 
another one, connected with the dusty filamentary structure close to
the outer planet orbit.  Its surface density is relatively low.  
The four rings 
outside of the outer planet orbit in the disk with 2 cm grains are almost 
equally spaced. Similarly to analogues rings in the disk with 1 cm grains, 
$\epsilon$ at the peak of the 
three most exterior dust rings
exceeds unity but the fourth one has a significantly lower dust surface density.

The dust surface distribution
in the viscous disk with 1 cm and 2 cm grains is overall similar to 
the one in the disk containing dust grains with $s_d=0.1$ cm.

When $s_{d}$ is taken to be 4 cm, and the viscosity in the disk is low, 
the dust substructures include the wide, but not very pronounced, ring 
outside the initial position of the outer planet, two very shallow gaps
filled in with the non regular dusty material, the void created behind 
the outer planet, the ring with the high dust surface density between planets
and no central cavity. 
After further 8000 orbits of the evolution, the dust surface density 
exhibits  five dust rings. 
Three of them with lower $\Sigma_{d}$ than in the case of $s_d$ = 2 cm
are generated at the exterior 
region of the orbit of the outer planet.
Two other dust rings formed in between the planetary orbits are similar 
to those of the $s_d$ = 2 cm case. 
One is very close to the orbit of the outer planet
along which a slight dust surface density enhancement relative to the gap
is present and another
narrow, high dust surface density ring is located roughly half way between
the orbits of the planets. 
In this case, the innermost region in the disk is not depleted as in the
cases of $s_d$ = 0.1 and 2 cm, but resembles rather the $s_d$ = 0.01 cm case
with $\Sigma_{d}$ close to its initial 
value. Moreover, there are substructures formed in this region such as two
shallow annular gaps. The void present after 10000 orbits grown in size.
The dust surface density in the viscous case is still very similar to
that described before for the $s_d=0.1$ and 2 cm cases. The only noticeable
difference is the presence of the void behind the planet.

The structure of the dusty environment formed during the two-super-Earth
migration in the viscous disks of gas and dust is less varied than the
one in the low-viscosity disks. The gaps
created by the planets are very shallow, so the contrast between the dust
density of the gap and the surrounding disk is quite low. The spiral arms 
induced
by the planets are present in the dust surface density distribution.

\subsection{The possibility of forming second-generation planets} 

In our calculations presented in Section~\ref{subsec:sden-dsden-evo},
we have found that in the case of two super-Earths migrating in the
nearly inviscid disk of gas and dust, 
the efficient accumulation of dust particles can occur.
The very low viscosity disks are 
preferable environments for this process to take place. There are two
pronounced features which are created, namely a narrow ring between the 
planets and a wide ring outside the orbit of the outer planet. 
The narrow ring between the planets is particularly prone to the
instability and in consequence to a collapse giving a birth of a planetesimal, 
a planetary core or
a planet. 
In \cite{2024MNRAS.528.6538P} the possibility of 
a sandwiched planet formation has been proposed, considering 
two sufficiently massive planets on the fixed orbits. 
We show that this scenario holds also in the case of the migrating planets.
Therefore, the efficient dust agglomeration in the system of two super-Earths
close to the 2:1 commensurability, investigated in our work, 
can result in the formation of a particular compact system consisting
of two planets in the vicinity of the 2:1 MMR with a third planet between them.

\begin{figure}[htbp]
\centering
\includegraphics[width=1.0\linewidth]{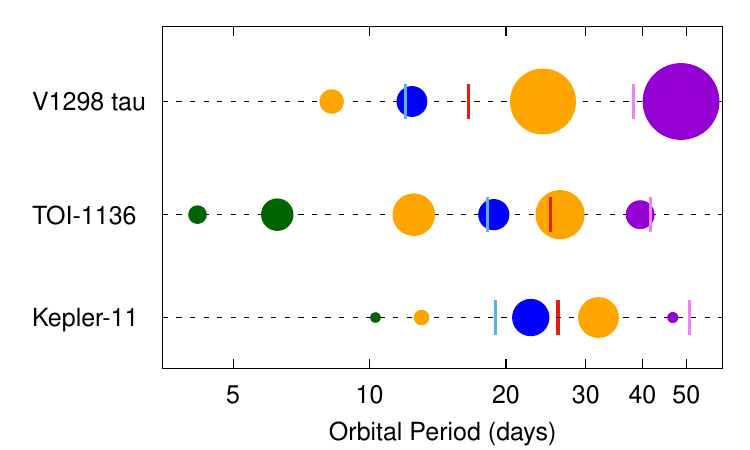}
\caption{The multi-planetary systems which contain two planets  outside 
the 2:1 MMR (orange) chosen to illustrate the possibility to form
two additional planets in the sandwiched (blue) and sequential (violet)
planet formation scenarios. The size of the circles 
represents the planetary mass obtained from NASA Exoplanet Archive. 
The red vertical lines indicate the position of the 2:1 MMR 
relative to the inner orange planet 
in each system.
The location of the dust rings between the two migrating super-Earths close to the 2:1
resonance, and those present outside the orbit of the outer super-Earth predicted from
our calculations are marked by the light blue and pink vertical lines in each system.
}
\label{fig:observation}
\end{figure}

We search the NASA Exoplanet Archive for the specific architectures, that
resemble the system considered in our simulations. We look for 
planetary systems containing three consecutive planets with the first 
(counting from the smallest to largest orbital periods)  
and the third one having the orbital separation ratio between 2 and 3 and
the mass of the first planet is lower than the mass of the third planet. 
Moreover, we exclude from our search those planets which orbital periods are 
less than five days to filter out the systems in which strong tidal
effects due to star-planet interactions are present.
We require also that the 
second planet is not the biggest among those three. 
We have found over a dozen of planetary systems which might satisfy 
our criteria, 
but only for a few of
those the masses are known sufficiently well to be able to verify the 
requirements. For this reason, in 
Figure~\ref{fig:observation} we show only three most interesting systems,
namely V1298 tau \citep{2026Natur.649..310L}, 
TOI-1136 \citep{2024AJ....167..70B} 
and Kepler-11 \citep{2013ApJ..770....131L}. 
The pairs of planets with the properties resembling those of the two 
super-Earths
migrating in the protoplanetary disks of gas and dust, considered in
this paper are marked in orange. The planets potentially set up through
the sandwiched planet formation are in blue and those potentially created 
in the process of the sequential planet formation are in violet. Planets
marked in green are the remaining planets in the systems. 
The Kepler-11 g is not included in this figure.
The red vertical lines in each system indicate the 
position of the 2:1 MMR relative to the inner orange planet. The location 
of the dust rings between the two migrating super-Earths close to the 2:1
resonance 
and those present outside the orbit of the outer super-Earth predicted from 
our calculations are marked by   
the light blue and pink vertical lines in each system. 
The size of each circle represents the planetary mass. 
An interesting outcome of our search is connected with the dust structure
formed on the exterior of the third planet in the system. We do not look
specifically for such configurations of four planets, but at least
for the systems presented here, the locations predicted from the calculations 
indicate the possible positions of the observed planets.  
Another intriguing fact based on our simulations can be noticed in the 
system TOI-1136. The period ratio of the outer super-Earth and the dust ring
between two planets in our calculations is approximately 1.4, which
correspond to the 7:5 resonance between TOI-1136 e and f in the observed
system. Our study indicate an interesting mechanism of the formation of this
second order resonance. 

The planets (blue) between two super-Earths (orange) could form from 
the accumulation of dust ($\epsilon > 1$) due to the presence of the pressure
maximum there as shown in Section~\ref{sec:two-planets-v5}.  
Similarly, the dust agglomeration exterior to the outer planet orbit can lead
also to the formation of a planet.
It should be stressed here that in this study we do not intend to
explain how  
those three systems formed, but rather to illustrate possible outcomes
of the orbital 
migration of two super-Earths close to the 2:1 resonance in the disk of 
gas and dust with different dust content and being subject to the resonance
overstability. 
In order to make an attempt to model the formation of planetary systems
adopting the sandwiched and sequential formation scenarios, it is necessary
to overcome the limitations of this investigation and we discussed it in the
next subsection.

\subsection{Limitations and outlooks}

In our two-fluid hydrodynamical simulations, we treat dust particles 
as a pressureless fluid. This assumption is appropriate when the 
dust-to-gas ratio $\epsilon$ is less than unity and the Stokes number 
of the dust grains is not too large, $St <1$. 
In our simulations with dust particle size $s_{d} \geq 1$ cm, one or 
more dust rings are formed and $\epsilon$ at the peak of the ring 
can exceed unity. Considering that the regions with $\epsilon > 1$ are 
quite narrow, we neglect the pressure effects caused by those dust 
rings on the numerical results. According to \cite{2009AA...502.385H} and
\cite{2017ApJ...844...142}
the particle fluid cannot be considered as pressureless when the Stokes 
number is larger than 0.5. In our calculations the Stokes number exceeds
this limit only at the initial position of the outer planet for the dust
grain size $s_d$ = 4 cm and is equal to 0.76. 
(see Appendix~\ref{sec:a3} for more details).
Our pressureless fluid set up may
therefore break down for this value of the Stokes number. 
We took the validity of the pressureless approximation into account while
presenting the finding of our calculations.
We intend to adopt the proper treatment of the dust pressure in the future
work.  

In this study, we consider the planet migration in the disks of gas and dust, 
assuming only the single-size dust particle presence 
in the disk. The size of the grains is kept unchanged during the  
simulation. 
The whole distribution of dust particle sizes should be considered 
and it is planned for the future investigations. Moreover, the dust  
coagulation and fragmentation are not considered in this work. 
The planet masses are fixed in the simulations, which means that the accretion
of the disk material on the planets is not included. 

Finally, we apply a locally isothermal disk in this work without considering 
the viscous heating and radiative cooling in the disk model 
\citep{2020AA...641A.125M}. 
The numerical resolution $1024\times700$ of our simulations is a result 
of adopting a large computation domain and being able to monitor the evolution
of two super-Earths for a long time (more than 20000 orbits). 
See Appendix~\ref{sec:a2} for more details).
Besides, the planets migration are calculated 
in two-dimensional hydrodynamical simulations and thus some mechanisms 
cannot be involved in the simulations such as dust settling, which may 
affect the numerical results. Therefore, more realistic modeling 
of the orbital evolution of multi-planetary systems should be performed
in three-dimensions. 

\section{Conclusions}
\label{sec:conclusion}

In this work, we have determined self-consistently 
the formation of dust 
substructures and the orbital migration of two super-Earths, which
triggered those substructures in a nearly inviscid locally isothermal 
disk of gas and dust. 
The planets with the masses of 5 M$_{\oplus}$ (inner planet) and 
10 M$_{\oplus}$ (outer planet), respectively, are placed close to the
position of the 2:1 resonance.
The dust component consists of the fixed uniformly sized grains and it is
treated as a pressureless fluid. The dust-feedback and dust diffusion are
taken into account in our 2D hydrodynamical two-fluid simulations.  
We have adopted five values of the size of the dust particles,
relevant for the observations in sub-millimeter and centimeter range 
of the electromagnetic spectrum, 
namely 0.01, 0.1, 1, 2 and 4 cm. We worked out how the size of the grains
affects the migration and the properties of the substructures induced by
the planets.     

Before reaching the resonance, the planets migrate inward convergently
and there are no significant differences in their migration rates
in the disks with and without dust grains. Also, the dependence on the
grain size is very weak. Therefore, in all cases, the planets are captured
in the 2:1 MMR.  
By design, the evolution of two resonant super-Earths in the gaseous disk 
exhibit
the libration overstability \citep{2024AA...686A.277A} and at some point 
planets leave the resonance.
We have established the impact of the dust on the duration of the 
resonance phase. In the disks with small grains (0.01 and 0.1 cm) 
the overstability proceeds 
in a similar way as in the purely gaseous disk. However,  
for 0.01 cm grains, the planets leave the 
MMR earlier than in the purely gaseous case, but later than in the disk
with 0.1 cm grains (Section~\ref{sec:two-planets-v5}, 
Figure~\ref{fig:evo-two-v5}). 
In the disks with large grains (1, 2 and 4 cm) the overstability 
is substantially modified 
and the duration of the resonance
is longer. In the case of the disk with 1 cm grains, the planets stay close
to the commensurability till the end of our simulations. If the 
grains are of 2 cm in size, the planets leave the resonance sooner, but not
as soon as in the disk with 4 cm grains, which take place at the similar
moment of time as in the purely gaseous disk.

After leaving the resonance the super-Earths in the disks with small grains
continue migrate inward faster than before entering the resonance and their 
relative migration is convergent. 
If the grains are large, planets leave the resonance migrating inward, but their
relative migration is divergent. 
We have shown that the divergent migration of planets is due to
their interactions with the dust ring formed between the planets. If
the dust-to-gas ratio, $\epsilon$, at the ring position is larger than 1 then 
the gas and dust 
torques are strongly affected by dust dynamics, resulting in faster 
migration of the inner planet and slower migration of the outer one 
(see Figure~\ref{fig:two-total-torque-v5}).
Moreover, the migration rate of the outer planet is influenced by the dust void
which is present in the  
co-orbital region of the planet in the disks with 2 cm and 4 cm grains 
(see Figure~\ref{fig:two-dust-torque-outer-sep}). 

The migration of the super-Earths and the substructures induced in the disk
by planets are coupled strongly with each other. 
The most robust substructures seen in our calculations  
are the rings and gaps in the dust surface density distribution.  
There are two specific locations where the rings,  
corresponding to the locations of the maxima of the pressure gradient 
in the disk, are generated
(Figure~\ref{fig:two-dsden-com-v5-1} and Figure~\ref{fig:two-dsden-com-v5-2}).
The narrow rings have their origin between the planet positions and their
density increases in time. Their shapes and the maxima of the density depend on
the size of the dust grains, but they remain narrow and for the large grains
(grains with their sizes larger or equal to 1 cm)
the dust-to-gas ratio exceeds 1 at the ring 
locations.
The wide rings have emerged exterior to the orbit of the
outer planet. They undergo the fragmentation leading to the development
of the multiple ring structures if the dust particles accumulated
there have their sizes larger or equal to 1 cm. 

We have found that the central cavities need more time to be formed 
than the rings.
At the time of 10000 orbits in our calculations, there is no single case
of such substructures. In the case of the disk with the 1 cm grains, its central
cavity is still not fully developed. After additional 8000 orbits the central
cavities are seen in the disks with 0.1, 1 and 2 cm dust grains 
(Figure~\ref{fig:two-contour-disk-v5}).
The asymmetric substructures are also present in our calculations. They are 
associated mainly with the dust material close to the planet position. 
In the case of the disk with 4 cm grains, the dust void is particularly pronounced.  

In this investigation, we have focused on the dust substructure formation in 
the nearly
inviscid disks of gas and dust induced by two migrating super-Earths. However,
for comparison, we have also considered disks with the
viscosity parameter equal to $10^{-3}$ (Section~\ref{sec:two-planets-v3}), 
two orders of magnitude higher than
in the nearly inviscid case. 
In the disks with 0.01, 0.1, 1 and 2 cm size grains, the planets enters 
into the 2:1 MMR, exhibit a limit cycle behavior as in a purely gaseous
disk \citep{2024AA...686A.277A} and stay in the resonance till the end 
of the simulations. In the disk with 4 cm grains, the planets arrive into 
the 2:1 MMR later, both eccentricities are lower than in the other cases 
(see Figure~\ref{fig:evo-two-v3}), the overstability is not present and 
the planet remain in the resonance till
the end of the computations.
In the viscous disks the substructure formation
is less efficient, which means that only a very shallow gap and a 
inconspicuous wide ring, located outside the outer planet orbit, 
are created 
(see Figure~\ref{fig:two-sden-dsden-v3}). The spirals excited by
the planets (see Figure~\ref{fig:two-contour-disk-v5}),  
which have not been seen in our calculations of the nearly inviscid 
disks, are present in such disks. 

Another interesting result has been obtained in the calculations of a single
super-Earth migrating in the disks of gas and dust 
(Section~\ref{sec:single-case}), performed with a 
motivation to facilitate the interpretation
of the simulations with two-migrating super-Earths, the main focus of our 
studies. 
The partial gap and dust rings are formed in the vicinity of the planet, 
see also \cite{2017ApJ...843..127D}. 
The location of the largest dust accumulation, which is outside 
the orbit of the planet with the mass of 10 M$_{\oplus}$ is close 
to the 7:5 resonance with the planet. The smallest distance to this
commensurability is obtained for the disk with the 1 cm dust grains.  
The exception is the case of the disk with the 0.01 cm grains, where 
the position of this ring is close to the 3:2 MMR. The small
accumulations, interior to the planetary orbit in this case, are close 
to the 1:3 and 2:3 
commensurability with the planet.
The resonance structure of the dense rings of
the 0.1 cm particles moving to larger radii in the presence of  the planet with 
the mass of 0.1 masses of Jupiter has been found also 
in \cite{2004A&A...425L.9P}. 
The co-orbital dust features, such as dust voids or  filamentary structures,
are observed 
in the disk with the moderate dust grains (1 cm), which is consistent 
with the previous works \citep{2018ApJ...855L..28B, 2025AA...698A..21C}.  

Our findings concentrate on the properties of the dust substructures seen
in the distribution of the dust surface density in the protoplanetary disks. 
The comparison
with the observed substructures requires further investigations and is not
the subject of this work. 
Instead, in this study, we have indicated another link with the 
observations, namely a link between the pronounced dust rings formed 
by the two migrating super-Earths, in the
two particular locations of the disk relative to the planets, and the observed
architectures of the known confirmed planetary systems. We have found 
in the NASA Exoplanet Archive a few
promising examples, one of those is of particular interest and it is TOI 1136
(see Figure~\ref{fig:observation}).
Further improvements in the protoplanetary disk modeling together with the
higher precision data from the new generation observational facilities like
upgraded ALMA, ngVLA, forthcoming SKA and the space mission 
PLATO will contribute enormously to the understanding of the connection 
between the early phases of planetary system formation and the architectures 
of the mature systems.

\acknowledgments

Z.C. gratefully acknowledge support by the National Natural Science Foundation of China (grants 1240030393).
Most of the simulations were performed on the High Performance Computing Resource in the Astronomy Department of Wuhan University, and Z.C. is indebted to Chang Shu for the continuous support on the technical maintenance of the servers.

%


\software{FARGO3D \citep{2019ApJS..241..25}}

\vskip12pt


\bibliography{zijia}{}
\bibliographystyle{aasjournal}





\appendix

\section{The steady-state profile of a disk of gas and dust with a fixed
particle size}
\label{sec:a1}

\begin{figure}[htbp]
\centering
\includegraphics[width=0.5\linewidth]{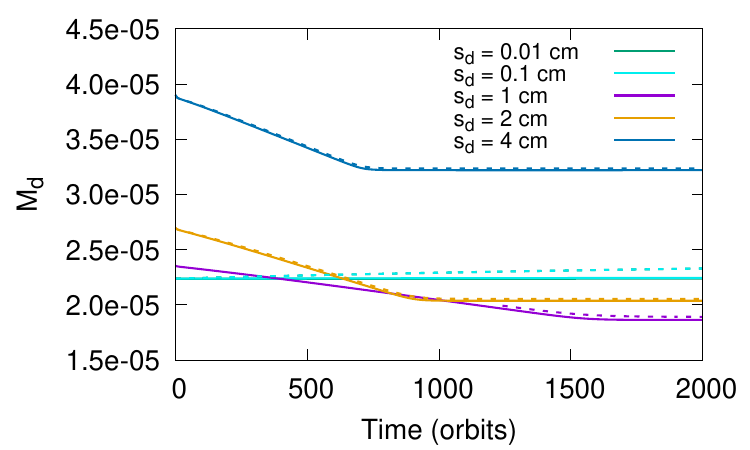}
\caption{The total dust mass in the computational domain of the simulations 
with different $s_{d}$ as a function of time. The solid and dashed lines 
indicate the results obtained with 
$\alpha_{0} = 10^{-5}$ and $10^{-3}$, respectively. The evolution
of $M_d$ for the disks with 0.01 and 0.1 cm grains is very similar, so the 
green and cyan lines practically overlap each other.}
\label{fig:empty-disk-dustmass}
\end{figure}

In our simulations, we consider uniformly sized dust particles.
To obtain the steady-state background solution, we run the simulations
of an empty disk (disk without planets), described in
subsection~\ref{subsec:physical-disk},
with different $s_{d}$ and
$\alpha_{0}$.
The achievement of the steady-state profile is illustrated
in Figure~\ref{fig:empty-disk-dustmass} showing
the total dust mass present in the computational domain of the
simulations, $M_{d}$, as a function of time. The results
for $\alpha_{0} = 10^{-5}$ and $10^{-3}$ are indicated by solid and dashed
lines, respectively.
In the case of disks with the large dust particles, namely $s_{d} \geq$ 1 cm, 
we find that $M_{d}$ decreases before the steady value is reached. 
The larger $s_{d}$, the faster the steady value of $M_{d}$ is achieved. 
In this case, the $M_{d}$ evolution is not very sensitive to the
value of $\alpha_{0}$ adopted in the simulations. 
In the case of disks with the smaller dust particles 
($s_{d} \leq$ 0.1 cm), $M_{d}$ continues to increase, but with a very low
rate. In this case, $M_{d}$ increases slightly faster if the viscosity 
is higher.
Based on these results, we conclude that $M_d$ contained in
the disks with
$s_{d} = $ 1, 2 and 4 cm has reached its steady-state in 2000 
orbits
and that the same statement is justified in the case of the disks with   
$s_{d} =$ 0.01 and 0.1 cm, taking into account the very slow increase of 
$M_{d}$ in time. Once the time needed to get the steady-state background
solution has been identified, we have verified that indeed the profiles
of $\Sigma_{g}$ and $\Sigma_{d}$ after 2000 orbits underwent 
only minimal alteration 
in the disks without embedded planets.

\section{The effects of the numerical resolution adopted in the simulations on 
the results}
\label{sec:a2}

The resolution applied in our simulations is taken to be $1024 \times 700$ 
in the azimuthal and radial directions, respectively, and on purpose it is 
the same 
as in \cite{2024AA...686A.277A}. 
Prior studies, with the aims to investigate the dust structure formation due to
the presence of a single super-Earth in the disk, have generally adopted 
a higher resolution in the calculations.
To achieve this, it was necessary to set a relatively small computational 
domain, which is suitable  
for a planet on the fixed orbit \citep{2018ApJ...855L..28B, 2025AA...698A..21C, 2017ApJ...843..127D, 2018ApJ...866..110D, 2025AA...694A.279R} or limit
a period of time during which the planet migration could be followed in 
the simulations \citep{2020MNRAS...497..2425H}. 
In this work, we have performed the simulation in a large computational
domain 
with $r \in [0.2, 7.0]$ to monitor the migration of two super-Earths in 
the period of time longer than 20000 orbits. Therefore, the cell size in 
the vicinity of the planets are not as small as applied in previous works. 

\begin{figure}[htbp]
\centering
\includegraphics[width=1.0\linewidth]{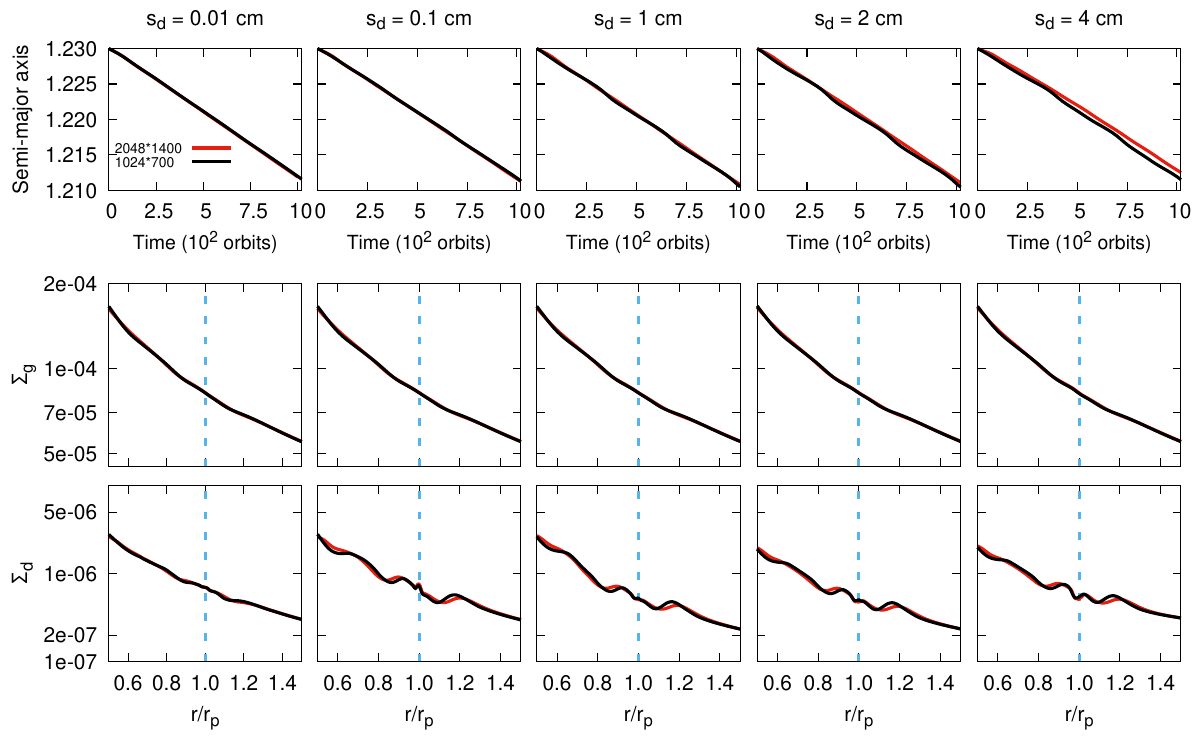}
\caption{From top to bottom: The evolution of the semi-major axis of a single 
planet with $q = 1.5 \times 10^{-5}$ and $r_{0} = 1.23$, the gas and dust 
surface density as a function of $r/r_{p}$ at t = 1000 orbits in the 
simulations with the resolution of $1024 \times 700$ and $2048 \times 1400$ 
for different $s_{d}$. The dashed vertical light blue line in the middle and 
the bottom panels indicates the position of the planet.}
\label{fig:resolution}
\end{figure}

In order to check how the numerical results might be affected by the 
adopted numerical resolution, we run the simulations for a single planet 
case with $q = 1.5 \times 10^{-5}$ and $r_{0} = 1.23$ increasing the  
resolution to $2048 \times 1400$ and compare the results with the 
simulations shown in Section~\ref{sec:single-case}. The evolution of the 
semi-major axis, the gas and dust surface density in the vicinity of the 
planet at t = 1000 orbits with different resolutions are presented 
in Figure~\ref{fig:resolution}. 
From the top panels, we can see that the migration rates of the planet 
in the disks with $s_{d}$ = 0.01 cm and 0.1 cm in two adopted
resolutions are similar to each other, while in the disks with 
$s_{d} =$ 1, 2 and 4 cm,   
the planet migrates a little bit slower when the resolution is higher.

The gas surface density profiles in the vicinity of the planets, 
shown in the middle panel, are not affected significantly by the
increase of the numerical resolutions adopted in the calculations. 
It holds independently of the dust content of the disks. 
The situation is a bit different with the dust surface density profiles
presented in the bottom panel. While the higher resolution does not 
influence noticeable the co-orbital regions of the planets, the edges of
the partial gaps are slightly reshaped
when $s_{d} \geq$ 0.1 cm.

In summary, it is unlikely that the results obtained in this work are
significantly affected by the numerical resolution adopted. 
We expect that our findings are robust, but the details of the
migration rates and/or dust substructure properties might be slightly changed.

\section{Changes of Stokes number at the planet's position during 
the disk evolution}
\label{sec:a3}

\begin{figure}[htbp]
\centering
\includegraphics[width=0.45\linewidth]{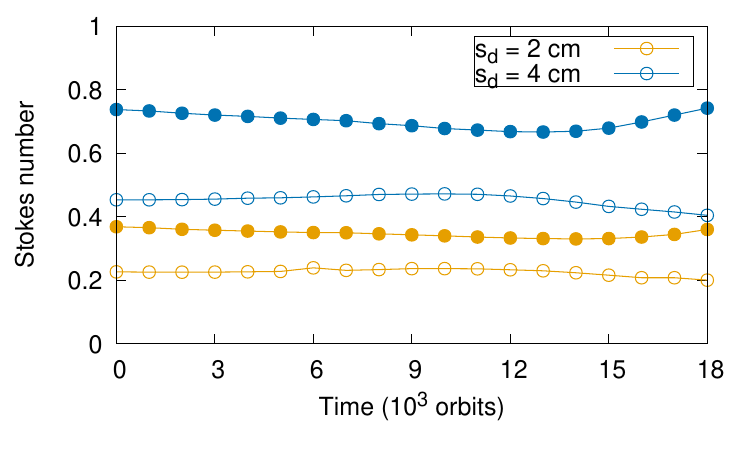}
\includegraphics[width=0.45\linewidth]{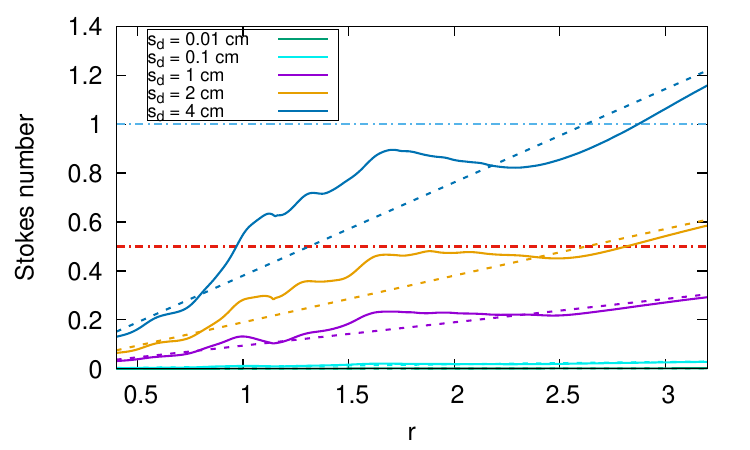}
\caption{Left: Stokes number at the location of two planets as a function 
of time in the simulations with $s_{d}$ = 2 cm and $s_{d}$ = 4 cm. 
The viscous parameter $\alpha = 10^{-5}$. The empty and solid circles denote 
the $St$ number at the positions of the inner planet and outer planet in each case, respectively. 
Right: Stokes number as a function of $r$ in the simulations with different 
$s_{d}$ at t = 0 and 18000 orbits, which are indicated by the dashed and 
solid lines, respectively. The red and light blue horizontal dashed-dotted lines represent $St = 0.5$ and 1, respectively.}
\label{fig:st-planet-evo}
\end{figure}

As stated in Section~\ref{sec:methods}, in our simulations, we consider 
the fixed dust particle size, which means that the $St$ number is not constant 
in the whole computational domain, but depends on $\Sigma_{g}$ according 
to Eq.~(\ref{eq:st-number}). During the evolution, both planets migrate and 
perturb the gas distribution in their vicinity. 
As a result, $\Sigma_{g}$ at the planet positions changes  
during the migration and thus also the $St$ number. 
To check, how well our assumption of the pressureless fluid 
is satisfied, 
we calculate the $St$ number at the positions of two planets 
at different times with $s_{d}$ = 2 cm and 4 cm for two super-Earths 
migrating in the  
disk with a very low viscosity. The results are presented in the left panel of 
Figure~\ref{fig:st-planet-evo}.
In both cases,  $St$ number is relatively close to its initial value, 
which means that the dust drag in the vicinity of planets can be treated 
according to the values given in Table~\ref{tab:st-num}. 

In the right panel of Figure~\ref{fig:st-planet-evo}, we show the $St$ number 
as a function of $r$ in the simulations of two migrating super-Earths in 
the disks with a very low viscosity and the grain sizes $s_{d}$, considered 
in this work. The dashed lines indicate the initial 
$St$ number in each case while the solid lines indicate the $St$ number 
calculated using the azimuthally averaged $\Sigma_{g}$ at t = 18000 orbits. 
We confirm that the $St$ numbers are below the unity in 
the relevant computational domain, namely $r \in [0.2, 2.6]$, in  all 
the simulations,  
and are lower than 0.5, in the same domain, in all cases except for the $s_{d} = 4$ 
cm one.

\end{document}